# A Programming Language Oriented Approach to Computability

Bachelorarbeit
der Philosophisch-naturwissenschaftlichen
Fakultät der Universität Bern

vorgelegt von

Aaron Karper

2013

Leiter der Arbeit:
Professor Dr. Thomas Strahm
Institut für Informatik und angewandte Mathematik

# Contents

# Chapter 1

# Introduction

Computability and complexity are in some way the basics of what computer science is about. Where algorithmics might answer what the currently best known solution to a problem is, computability theory says if a solution can exist and complexity theory gives the problem (not the concrete algorithm!) a general assessment of its difficulty, i.e. no algorithm can solve the problem faster than its inert complexity.

For this, one needs an idea what a problem is, and when it is solvable. A problem can be understood as getting some kind of input data $x \in I_A$ and needing to produce an output value $y \in O_A$, in other words: computing a function $f : I_A \to O_A$[1].

A problem is then considered solvable, if there is in some way a solution to it, but what a solution looks like, divides the minds:

**Turing and his machines** Turing's approach is a very mechanical one: A machine that has memory in the form of an arbitrarily large tape and is in a certain state. Depending on the tape's content at the current position and the current state, it can write to the tape, change its state and move one cell to the left or to the right.

**Kleene composes functions** Kleene held it intuitively computable that there are some simple functions such as the projections, the constant functions, and the successor are computable and that the composition of two computable functions is again computable. Finally every function can be coded and the codes can then be executed.

**Markov rewrites strings** Markov noted that executing an algorithm is basically nothing but rewriting a string (the memory) according to some predefined rules.

**Programming languages** While each of these approaches is interesting in their own right, they do not reflect the intuitions of a modern computer scientist, who already knows about programming languages and is closer to them then to building machines or even mathematical constructs.

Nearly all modern languages try to capture the full spectrum of solvable problems. For the purposes of this text, the `WHILE` language will serve as a minimal coding for computability.

### 1.0.1 What is a programming language?

We have an intuitive notion of what a programming language is, which probably goes like "A language, in which programs can be written" or "An executable language". But what makes a string of letters executable?

---

[1] Equivalently, a problem can be characterized as deciding if $x \in^? B$.

The difference is just that we know how a programming language should be interpreted, i.e. a program can be mapped to a function and thus in the realm of mathematics.

**Definition 1.** A *partial function* from $A$ to $B$ is a function
$$f : A \to B_\perp = B \cup \{\perp\}$$
where $\perp$ signifies that no value exists for that input.

For example $[\![A]\!](x) = \perp$ means, that the program $A$ does either give an error or does not terminate on the input $x$.

**Definition 2.** A *semantic function* $[\![.]\!]_A$ for a programming language $A$ is a function
$$[\![.]\!]_A : A \to (I_A \to O_A \cup \{\perp\})$$
where $\perp$ signifies either an error or that the function does not return. Then the function $[\![.]\!]_A$ takes a valid $A$ program and gives a function that maps inputs for $A$ programs to outputs (if any).

The semantic function defines what a program *means*, while its syntax defines *how it should look like*. It typically uses other functions to denote parts of the program.

**Example.** As a part of a programming language, we want to know the semantics of `572`. Note that `572` is a string, not the number. However $[\![572]\!]_{dec} = 5 \cdot 100 + 7 \cdot 10 + 2$ *is* the number, not the string.

When describing a programming language, the terms *expression* and *statement* will be used often. An *expression* is a representation of data used in the language. It returns a value and can depend on the state of the computation, e.g. to evaluate variables. Some expressions even change the state of the computation. A *statement* does not return a value and just changes the state.

**Example.** In the assignment in a `C`-like language `int four = 2+2;`, the whole line is a statement since it introduces a new variable name, that can be referenced. `2+2` is an expression that returns 4. After this line, `four` is an expression as well and will return the current value of the variable.

## 1.1 The FOR language

$\langle expression \rangle ::=$ 'nil'
$\quad | \quad$ 'cons' $\langle expression \rangle\ \langle expression \rangle$
$\quad | \quad$ 'hd' $\langle expression \rangle$
$\quad | \quad$ 'tl' $\langle expression \rangle$
$\quad | \quad$ ':' $\langle symbol\text{-}name \rangle$
$\quad | \quad \langle expression \rangle$ '=' $\langle expression \rangle$
$\quad | \quad \langle variable \rangle$

$\langle statement\text{-}list \rangle ::= \langle statement \rangle$ ;
$\quad | \quad \langle statement \rangle$ ; $\langle statement\text{-}list \rangle$

$\langle block \rangle ::=$ '{' $\langle statement\text{-}list \rangle$ '}'

$\langle statement \rangle ::= \langle variable \rangle$ ':=' $\langle expression \rangle$
$\quad | \quad$ 'if' $\langle expression \rangle\ \langle block \rangle\ \langle else\text{-}block \rangle$
$\quad | \quad$ 'for' $\langle variable \rangle$ 'in' $\langle expression \rangle\ \langle block \rangle$

$\langle else\text{-}block \rangle ::= \langle empty \rangle$
$\quad | \quad$ 'else' $\langle block \rangle$

$\langle program \rangle ::= \langle name \rangle$ 'read' $\langle variable \rangle\ \langle block \rangle$ 'write' $\langle variable \rangle$

Table 1.1: The FOR syntax

### 1.1.1 The Elements

The FOR language contains only very basic commands, but they can be combined to implement a huge number of algorithms. As the data structure, we choose the humble *ConsCell*, that contains only a reference as the head and another as the tail[2]. Because the head and the tail point to *something*, we also need some atomic data, that it can point to. We introduce *Nil*, which points at nothing and is the basis for the *ConsCell* data. As a short cut notation, let

$$[a_1, a_2, \ldots, a_n] := ConsCell(a_1, ConsCell(a_2, \ldots ConsCell(a_n, Nil)\ldots)).$$

To make the programs easier to decipher, we also use named symbols, which are not strictly necessary: we could also use lists to code them.

The formal definitions of the semantics of FOR can be seen in the tables 1.2 and 1.3, but to gain familiarity with semantic functions, the following sections will explain the meaning of the definitions.

---

[2] An observant reader will notice that this structure stems from the building of linked lists.

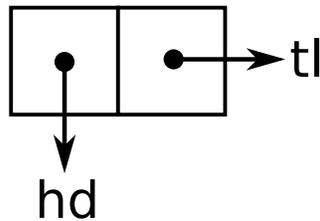

Figure 1.1: The *ConsCell*

In the `FOR` language, when we write the expression `cons` $e_1$ $e_2$, it will give us such a *ConsCell* with the evaluation of $e_1$ in the head and the evaluation of $e_2$ in the tail. The expression `hd` $e$ yields the head of the evaluation of $e$ and `tl` $e$ its tail. The expression `nil` just gives back $Nil$[3]. We also want to refer to stored values, so an identifier (e.g. `X`) can also be an expression, the evaluation of which depends on the current variable bindings.

There are very few types of statements in `FOR`, just three to be precise: The first is the *assignment* `X` $:= e$, which assigns the variable `X` the evaluation of $e$. Later, when `X` is used in an expression, it will reproduce this value. The second is the classic `if` statement, that only executes its first block, if the expression does *not* return `nil` and the else-block – if any – otherwise. Finally, there is the eponymous `for` $I$ `in` $e$ *block* loop, that works as follows:

1. On first entering the loop, the expression $e$ is evaluated and stored.

2. If the evaluation is $Nil$, we are done.

3. Otherwise the block is evaluated with the head of $e$ bound to the variable name $I$.

4. Then this procedure is run again with the tail of $e$.

### 1.1.2 Coding `FOR` programs

Some of the more interesting programs in the `FOR` language use other `FOR` programs as their input, for example a program *composition* that takes a program $p$ and another program $q$, such that $[\![[\![composition]\!](p,q)]\!](x) = [\![p]\!]([\![q]\!](x))$. In order to create such a program, we need some interpretation [**This would be a good point to start talking of functors, but then again, nobody wants to hear that in a practical text**] $\ulcorner . \urcorner : \texttt{FOR} \to I_{\texttt{FOR}}$.

One possible way to go about this is interpreting blocks and expressions as lists and denote them:

---

[3]Note the difference: `nil` is an expression in the language, literally the string `nil`, but $Nil$ is the mathematical entity. Similarly you should not mix up `cons` $a$ $b$ and $ConsCell(x,y)$

If $c = \{(name_i, value_i)\}$ is the variable binding at the time of the evaluation:

$$[\![\texttt{nil}]\!](c) = Nil \tag{1.1}$$

$$[\![:a]\!](c) = symbol_a \tag{1.2}$$

$$[\![\texttt{cons } A\ B]\!](c) = [\![(A.B)]\!] = ConsCell([\![A]\!](c), [\![B]\!](c)) \tag{1.3}$$

$$[\![\texttt{hd } A]\!](c) = \begin{cases} x, & \text{if } [\![A]\!](c) = ConsCell(x,y) \\ \bot, & \text{otherwise} \end{cases} \tag{1.4}$$

$$[\![\texttt{tl } A]\!](c) = \begin{cases} y, & \text{if } [\![A]\!](c) = ConsCell(x,y) \\ \bot, & \text{otherwise} \end{cases} \tag{1.5}$$

$$[\![A = B]\!](c) = \begin{cases} ConsCell(Nil, Nil), & \text{if } [\![A]\!](c) = [\![B]\!](c) \\ Nil, & \text{otherwise} \end{cases} \tag{1.6}$$

$$[\![name]\!](c) = \begin{cases} x, & \text{if } (name, x) \in c \\ Nil, & \text{otherwise} \end{cases} \tag{1.7}$$

Table 1.2: Semantics of FOR expressions

If $c = \{(name_i, value_i)\}$ is the variable binding at the time of the evaluation:

$$[\![name := a]\!](c) = (c \setminus \{(name, x)\}) \cup (name, [\![a]\!](c))$$

$$[\![\texttt{if } a\ block_1\ \texttt{else } block_2]\!](c) = \begin{cases} [\![block_2]\!](c), & \text{if } [\![a]\!](c) = Nil \\ [\![block_1]\!](c), & \text{otherwise} \end{cases}$$

$$[\![\texttt{if } a\ block]\!](c) = \begin{cases} c, & \text{if } [\![a]\!](c) = Nil \\ [\![block]\!](c), & \text{otherwise} \end{cases}$$

$$[\![\texttt{for } a \texttt{ in } v\ \{\ b\ \}]\!](c) = \begin{cases} c, & \text{if } [\![v]\!](c) = Nil \\ ([\![\texttt{for } a \texttt{ in } [\![\texttt{tl } v]\!](c)\ \{\ b\ \}]\!] \circ [\![b]\!] \circ \\ \quad [\![a := \texttt{hd } v]\!])(c), & \text{otherwise} \end{cases}$$

Table 1.3: Semantics of FOR Statements

$$\ulcorner \texttt{nil} \urcorner = symbol_{nilexp}$$
$$\ulcorner :a \urcorner = [symbol_{symbol}, symbol_a]$$
$$\ulcorner \texttt{cons a b} \urcorner = [symbol_{cons}, \ulcorner a \urcorner, \ulcorner b \urcorner]$$
$$\ulcorner \texttt{hd } a \urcorner = [symbol_{hd}, \ulcorner a \urcorner]$$
$$\ulcorner \texttt{tl } a \urcorner = [symbol_{tl}, \ulcorner a \urcorner]$$
$$\ulcorner variable \urcorner = [symbol_{var}, symbol_{variable}]$$
$$\ulcorner a := e \urcorner = [symbol_{assign}, symbol_a, \ulcorner e \urcorner]$$
$$\ulcorner \{b_1;b_2;\ldots;b_n\} \urcorner = [\ulcorner b_1 \urcorner, \ulcorner b_2 \urcorner, \ldots, \ulcorner b_n \urcorner]$$
$$\ulcorner \texttt{for } a \texttt{ in } e\ block \urcorner = [symbol_{for}, symbol_a, \ulcorner e \urcorner, \ulcorner block \urcorner]$$
$$\ulcorner \texttt{if } e\ if\text{-}block\ else\text{-}block \urcorner = [symbol_{if}, \ulcorner e \urcorner, \ulcorner if\text{-}block \urcorner, \ulcorner else\text{-}block \urcorner]$$

Table 1.4: Coding the `FOR` language in its own data

### 1.1.3 `FOR` computability

The probably most important property of the `FOR` language is that programs in it always terminate. The reason for this is, that upon entering the for loop, the number of repetitions is fixed as the length of the evaluated expression. Since we can't build an infinite expression in the finite time before the loop, the program terminates.

When experimenting with the language, one quickly finds, that many important functions are `FOR` computable:

- Constant functions.
- Addition.
- Multiplication – as repeated addition.
- Exponentiation – as repeated multiplication.
- Unary/binary conversion.
- Testing if a given number is prime.
- ...

As we can see, it is relatively simple to generate huge numbers using the `FOR` language and then generate lists with that length.

At first glance it seems that any computation can be defined this way, but sadly, that is not the case.

Intuitively the interpretation of a `FOR` program should be computable. We can do it in an algorithmic way and so it is only reasonable to expect `FOR` to be able to interpret itself. As we will see, that can not be the case:

**Theorem 1.** *There is no* `FOR` *program exec that takes* $ConsCell(program, input)$ *as its argument and returns* $[\![program]\!](input)$.

*Proof.* Assume, there was a procedure in `FOR` *exec* that takes $(program.input)$ as its argument. Now the following procedure would surely be a `FOR` program too:

```
inverse read X {
  result := [exec](X.X)
  if result {
    Y := FALSE;
  } else {
    Y := TRUE;
  }
} write Y
```

We have

$$[\![inverse]\!](program) = \begin{cases} \text{TRUE}, & \text{if } [\![program]\!](program) = \text{FALSE} \\ \text{FALSE}, & \text{otherwise} \end{cases}$$

It outputs `TRUE` iff the given program outputs `FALSE`. What then is $[\![inverse]\!](inverse)$? Assume first that it is `TRUE`, then by definition it is `FALSE` – and vice versa! So it neither returns `TRUE` nor `FALSE`. The only way that would work would be if it didn't return anything at all, but as we have seen, all `FOR` programs terminate in finite time, so that can not be the case. Therefore `inverse` can not be a `FOR` program and by extension `exec` is not a `FOR` program.[4] □

This is unfortunate, not only because we have seen that there are functions that are not `FOR` computable, but that in general, any language that supports the building blocks of `inverse` either doesn't support self-interpretation or it contains programs that will not halt.

As we try to capture *all* intuitively computable functions, it is not acceptable to leave self-interpretation out, so in the next chapter, we will explore a language that will contain non-halting programs.

### 1.1.4 The `FOR` in real programming languages

It turns out, that while the `for` keyword exists in most languages in one form or another, few acurately model the `FOR` languages intention. Foremost most languages cheat around the property of `FOR` loops always terminating: In C-like languages, the loop variable is not immutable and therefore

---

[4]This is the quitessential uncomputability proof: A coding of the function is given that can be evaluated, but running the program on itself leads to complications.

```
for (int i = 0; i < 10; i++) {
  // ...
}
```

might very well not terminate, if `i` is decremented in the body. Iterator-based loops can be tricked by implementing an iterator that does not halt, i.e. produces a new value on any `next`.

### 1.1.5 The FOR in mathematics and logic

The FOR computable functions match a category of functions known as *primitive recursive*. A function is *primitive recursive*, if it is either

1. A successor function $S_i(x_1, \ldots, x_n) = x_i + 1$

2. A projection function $P_i(x_1, \ldots, x_n) = x_i$

3. In the form of primitive recursion:

$$f(n, x_1, \ldots, x_n) = \begin{cases} g(x_1, \ldots, x_n), & \text{if } n = 0 \\ h(f(n-1, x_1, \ldots, x_n), x_1, \ldots, x_n), & \text{else} \end{cases}$$

Where $g$ and $h$ are primitive recursive.

It is easy to see that FOR can compute any primitive recursive function: The successor function and the projections are trivial, for the primitive recursion, we have programs $[\![G]\!] = g$ and $[\![H]\!] = h$ by induction assumption.

```
F read NX {
  N := hd NX
  Xs := tl NX
  Repetitions := [unary](N)
  Result := [G](Xs)
  Counter := 0
  for I in Repetitions {
    Result := [H](Result.Xs)
  }
} write Result
```

**Exercise 1**   FOR **computable functions are primitive recursive**

1. Can you model an `if` statement in primitive recursion?
2. How would you implement a `for` loop with primitive recursion?
3. How could we model assignment in the primitive recursive functions?
4. *Difficult:* Show that *ConsCell* can be modelled in the natural numbers, with primitive recursive functions handling `cons`, `hd` and `tl`.

## 1.2 The WHILE Language

In the last chapter we saw an example of a function that is not computable with a FOR program[5]. However with a simple addition to the language, we gain all we need for a language.

The new statement is called while $e$ {$block$}, and it does what one would expect:

- It takes an expression $e$ and a block *block*.

- If the evaluation of the expression yields nil, it does nothing.

- Otherwise, it executes the block and repeats this procedure.

$$[\![\texttt{while}\, e\, block]\!](c) = \begin{cases} c, & \text{if } [\![e]\!] = Nil \\ ([\![\texttt{while}\, e\, block]\!] \circ [\![block]\!])(c), & \text{otherwise} \end{cases} \quad (1.8)$$

This new statement does not necessarily terminate, in fact WHILE (nil.nil) {} would never halt. This means, that the semantic function does not give a total function back – for some inputs the interpretation of the source does not halt.

Looking back at the proof of the uncomputable function in FOR called *inverse*, that took a program and returned the boolean inverse of that programs output when run with itself as input. We asked what $[\![inverse]\!](inverse)$ would be. Since it can't be $TRUE$ nor $FALSE$, it must be $\bot$. This can also be seen in that $[\![eval]\!](inverse.inverse)$ is just an infinite recursion.

Of course it could be that while this proof does not work, something else might prevent us from implementing the eval procedure. As we will see in 2.3, that is not the case.

Part of this work is an implementation of the WHILE programming language, which can be found at https://github.com/zombiecalypse/Bachelor-Thesis/wiki.

---

[5] The semantic function of FOR

# Chapter 2
# Computability

## 2.1 Language Transforms

In this chapter, we'll discuss how we can use the notion of a data representation of a program to get a standard toolchain.

### 2.1.1 Language Subsets

**Definition 3.** Let `A` and `B` be two languages such that each valid `B` program is also a valid `A` program. Further $\forall b \in B \, \forall d \in I_B : [\![b]\!]_A (d) = [\![b]\!]_B (d)$

Then $B$ is a language subset of $A$ (and conversely $A$ is a superset of $B$), writen $B \subset A$.

**C++ and C**  C++ was designed to be a object-oriented superset to the popular `C` programming language, so that the new `C++` code could use legacy `C` code without modification. This notion was important to raise the acceptance of `C++` with programmers and eased switching.

### 2.1.2 Interpreter

Today programmers and machines seldom speak the same language. Programming in machine language is difficult, error-prone and unportable to name only a few drawbacks. However it seems reasonable to expect to be able to execute ones code nonetheless. Typically, we want our computer to interpret what we mean in our high-level programming language. An *interpreter* is such a program.

**Definition 4.** An *interpreter* of $A$ is a program *interp* such that

$$[\![interp]\!]_P : A \times I_A \longrightarrow O_A$$
$$[\![interp]\!]_P(a, d) = [\![a]\!]_A (d)$$

Since parsing is not part of this text, we will assume, that a convenient format is already given, like the one introduced in 1.1.2.[1]

Interpreters are typically the first step in implementing a language: they are relatively easy to write and therefore allow experimenting. The downside is that working on the AST and constantly translating typically takes longer than an equivalent $A$ program would take. This *overhead* is often just a fixed factor, but that factor could be 100, making the interpreted program a hundred times slower than the native one.

### 2.1.3 Compiler

Given the problem of a language that can not be executed directly, there is also another approach that can be taken instead of interpretation: We could translate the program into a native one in the executable language. This process is called *compiling* and the program that does this is a *compiler*.

---

[1]If you are interested in the whole story, [1] offers a good introduction into the theory of building tools for a new language.

**Definition 5.** A *compiler* for the language $A$ into the language $B$ is a program $compile \in P$, such that

$$[\![compile]\!]_P : A \longrightarrow B$$
$$[\![[\![compile]\!]_P(a)]\!]_B = [\![a]\!]_A$$

Normally $P = B$, but if it is *not*, then *compile* is a so-called *cross-compiler*.

What does it mean, if we can express an compiler for $A$ in the language $B$? It means, that we can solve any problem in $B$ that can be solved in $A$, by compiling the $A$ solution to $B$. This allows us to classify languages in the following way.

**Definition 6.** The language $A$ is *at least as powerful* as $B$, if there is an compiler $compile_{B \to A} \in A$. We write $A \leq B$.

Similarly, $A$ are *equally powerful* or *Turing equivalent* if $A$ can compile $B$ and vice versa. We write $A \equiv B$.

**Theorem 2.**   1. $\leq$ *is reflexive, that is* $A \leq A$.

2. $\leq$ *is transitive, that is if* $A \leq B \leq C$, *then* $A \leq C$.

*Proof.*   1. The compiler is the identity.

2. 
$$compile_{A \to B}^C = [\![compile_{B \to C}]\!]_A (compile_{A \to B})$$
and then we can compose $[\![compile_{B \to C}]\!]_C \circ [\![compile_{A \to B}^C]\!]_C$

$\square$

**How a modern compiler works**

A compiler typically has three stages, a frontend, a middle, and a backend.

The *frontend* transforms the language into a handy format, that is not necessarily similar to the input language. Compiler collections may have many frontends, that transform all kind of languages into this intermediate format.

The *middle* makes all kind of transformations on the immediate format, for example optimizations, type checking, .... This is where productive work can be done, so much work is typically invested here[2].

In the *backend*, the actual output is generated by transforming the immediate format to the output language. Separating this from the middle makes it easier to work for different output languages.

---

[2]See [1]

**How the `gcc` is ported** When a new machine architecture is build, there is the problem that there is not yet any compiler for it. The naïve solution would be to write a complete compiler in the new machine language, but that would be very cumbersome and inefficient to do that for every new processor build.

There are two parts to the problem: on one hand, there is not any compiler, that outputs the new machine lanugage and on the other hand, there is no compiler that runs on the new machines.

For the first problem, we see that of the three stages of a modern compiler, only the backend really depends on the output language. Often the backend has a general variant that can be parametrized for many architectures. [3]

The second problem is nowadays solved by cross-compiling – when the language the compiler is executed in and the output language differ. Another approach was to have a minimal (non-optimizing) compiler or an interpreter to do the first translation.

### 2.1.4 Specializer

While programmers generally know the notions of a compiler and possibly of an interpreter, the *specializer* is less well known. *Specialization* is the process of fixing a certain value in the source code, even though it was written in a way that would have allowed different values.

**Definition 7.** A *specializer spec* for the language $A$ is a program

$$[\![spec]\!]_P : A \times I_A \to A$$
$$[\![[\![spec]\!]_P (a, x)]\!]_A (y) = [\![a]\!]_A (x.y)$$

Informally speaking, the specializer moves an argument from the runtime to the compile time.

Typically, this is part of the optimizations that occur during compilation. For example, if we encountered the call `fib(n, true)` for the function

```
int fib(int n, bool debug) {
  if (debug) printf("Call with %d", n);

  if (n == 0 || n == 1) return 0;

  return fib(n-1) + fib(n-2);
}
```

then the specializer might remove the first test and just leave the `printf`, which is more efficient and removes the "loose end" `debug` from the runtime.

For a specializer to work, it has to prove, that a certain part of the program only depends on the given value and static data and then evaluate that. This sounds easier than it is:

---
[3] For the whole process of writing a `gcc` backend, see [9]

**Example.** Given the expression `a+b+1`, where `a` is statically known to be 5, we know that it could be specialized to `b+6`, but if we interpreted it as `(a+b)+1` *or* `a+(b+1)`, no subexpression would be independent of `b`.

For a deeper look at the applications and implementations of specializers, see [7].

**Futamura Projections**

The notion of a specializer as a transformer of source code has lead to some interesting observations by Yoshihiko Futamura [4], which are now known as the Futamura Projections:

1.  
$$[\![[\![spec]\!]_P (interpret_A, source)]\!]_P (inp) = [\![interpret_A]\!]_P (source.inp)$$
$$= [\![source]\!]_A (inp)$$

    So we can get an executable, if we specialize the interpreter with the source of our program.

2.  
$$compiler_{A \to P} = [\![spec]\!]_P (spec, interpret_A)$$
$$[\![[\![spec]\!]_P (spec, interpret_A)]\!]_P (source) = [\![spec]\!]_P (interpret_A.source)$$

    Therefore we can get a compiler, if we specialize the specializer with the interpreter.

3.  
$$[\![[\![spec]\!]_P (spec.spec)]\!]_P (interpret_A) = [\![spec]\!]_P (spec.interpret_A)$$

    So we can get a program, that takes an interpreter for any language $A$ and produces a compiler for $A$ from it.

This approach would make it as easy to generate a compiler as it is to program an interpreter, so why are not all compilers generated this way?

As discussed, finding specializable parts of the interpreter is not as easy as the equations make it look, so it is not surprising, that a correct specializer will not catch every possible optimization. In fact, optimizing the intermediate format allows many other optimizations besides the specialization and so a pragmatic compiler writer will prefer to do such things by hand, instead of enhancing the specializer.

---

[4] [4]

### The PyPy project

**Example.** The PyPy project is an attempt to implement the popular Python[5] itself in a subset of Python (called RPython). Since Python is an interpreted language, it would seem that this approach would lead to very slow execution, but that is not the case: PyPy uses Just-In-Time (JIT) specialization and compilation techniques in part described in [10].

While the approach described in 2.1.4 is understood to be executed before the actual program is run, it is also possible to run it in parallel to the actual computation: Now the specializer can use statistical information on the values. For example, while it might not be obvious from the source that a certain value is constant and therefore a static specializer might fail to set in, but a dynamic specializer can determine this and produce a specialized function to call.

For a highly dynamic language like Python, it can lead to a hundredfold speedup for very repetitive arithmetics[6].

### Theoretical Results

Proposing that we can statically fix arguments in our program leads important results:

**Theorem 3.** *If there is an interpreter $interpret_B \in A$ then $B \leq A$.*

*Proof.* By the second Futamura projection

$$compile_{B \to A} = [\![spec]\!]_A \, (spec, interpret_B)$$

□

Note however that the reversal is not necessarily true: For example, we know, that FOR is has no $interpret_{\text{FOR}}$, but FOR $\leq$ FOR still holds. We will see later that a self-interpreter allows us to identify the two notions.

---

[5]http://python.org/
[6] [10]

## 2.2 Turing-Completeness of a Language

In the dawn of computer science there were serveral ideas what computable means. Does it mean, that a machine can compute it, as Turing suggested? Or should we define some functions that should be computable and how they can be combined to discuss the computability of a problem?

As it turned out, it does not matter – the problems that can be solved in each of them are the same. This was shown by proving that the different mechanisms can simulate each other. By extension, a new formulism only needs to be able to simulate any of the existing formalisms to be at least as powerful as any of them. Thus came the expression *Turing complete* to describe, that a certain formalism can build any Turing machine.

This chapter is build as exercises, because solving the problems gives a good impression, how one could go about proving other things to be turing complete.

### 2.2.1 The Turing Machine

The Turing Machine (`TM`) is a formalism that focuses on a possible physical implementation of a computing machine, although far from what is used today. It features a potentially infinite amount of tape on which the machine operates and which it uses as storage. From this tape, the machine can read and write the current cell and move its read-write-head one cell to the left or the right. Finally, the machine has a finite number of "states of mind", comparable to a finite state automaton.

In order to formally analyse the `TM`, we can code what such a machine would do in the following way:

**Definition 8.** A Turing Machine $M$ is a tuple $(Q, \Sigma, \Gamma, \delta, q_0, q_{accept}, q_{reject})$ where

- $Q$ are the states of mind the machines can be in.

- $\Gamma$ is the tape alphabet, that can be used during the computation.

- $\Sigma \subset \Gamma$ is the input alphabet, of which the initial tape is a word.

- $\delta : Q \times \Gamma \to Q \times \Gamma \times \{L, N, R\}$ is the transition function, which codes, how the `TM` reacts to a state and the current symbol – i.e. what will be the new state of mind, what symbol will be written and in which direction the head will move.

- $q_0 \in Q$ gives the initial state.

- $q_{accept}, q_{reject} \in Q$ signify that the machine has stopped – successfully or not.

To run a `TM` means to give it a word $w \in \Sigma^*$, place the head at some cell on the tape and then execute the transitions indicated by the transition function until we get into any of the states $q_{accept}$ or $q_{reject}$.

**Definition 9.** A *configuration* of a Turing machine $M = (Q, \Sigma, \Gamma, \delta, q_0, q_{accept}, q_{reject})$ is then a tuple $c = (tape, s) = (l, h, r, s) \in \Gamma^* \times \Gamma \times \Gamma^* \times Q$. This allows us to define a step that a TM makes.

$$step(tape, s, (\tilde{h}, \tilde{s}, d)) = \begin{cases} (tape, s), & \text{if } s \in \{q_{accept}, q_{reject}\} \\ (l, \tilde{h}, r, \tilde{s}), & \text{if } d = N \\ (\tilde{h}\, l, r_1, r_{2,\ldots}, \tilde{s}), & \text{if } d = R \\ (l_{2,\ldots}, l_1, \tilde{h}\, r), \tilde{s}), & \text{if } d = L \end{cases}$$

or in other words: if the head moves *right*, then the rewritten head is pushed to the left stack and the new head is popped off the right stack.

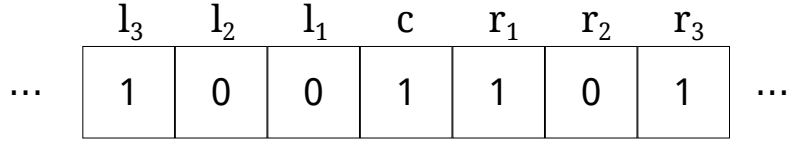

Figure 2.1: Tape of the Turing machine

We write $c_1 \vdash_M c_2$ to denote that $step(c_1, \delta(c_1)) = c_2$ and say that one *step* of $M$ transforms $c_1$ to $c_2$.

Now $c_1 \vdash_M^* c_2$ means that an arbitrary number of steps lead $c_1$ to $c_2$.

$$[\![M]\!]_{\text{TM}}(tape_1) = \begin{cases} tape_2, & \text{if } (tape_1, q_0) \vdash_M^* (tape_2, q_{accept}) \\ \bot, & \text{if there is no such } tape_2 \end{cases}$$

**Example** (Adding two binary numbers). When we implement binary addition, we first need to think about how to represent the problem. One of the easiest way to do that is to use 0 and 1 to represent the digits, + to separate the arguments, = ending the tape and assume that the numbers start with the least significant number on the left (which is the reverse of the usual notation, but easier to calculate with).

1: Start with *carry* $= 0$ in mind.
2: **if** tape shows 0 **then**
3:     keep same *carry* in mind
4: **else if** tape shows 1 **then**
5:     **if** *carry* $= 0$ **then**
6:         have *carry* $= 1$ in mind
7:     **else if** *carry* $= 1$ **then**
8:         have *carry* $= 2$ in mind
9: **else if** tape shows + **then**

```
10:      keep same carry in mind
11:      have finished = 1 in mind
12: Move to the right until you hit +.
13: if tape shows 0 then
14:      keep same carry in mind
15: else if tape shows 1 then
16:      if carry = 0 then
17:          have carry = 1 in mind
18:      else if carry = 1 then
19:          have carry = 2 in mind
20:      else if carry = 2 then
21:          have carry = 3 in mind
22: else if tape shows + then
23:      keep same carry in mind
24:      if finished = 1 then
25:          have finished = 2 in mind
26: Move to the right until you hit =.
27: Move to the right until you hit #.
28: if carry = 0 then
29:      write 0
30:      have carry = 0 in mind
31: else if carry = 1 then
32:      write 1
33:      have carry = 0 in mind
34: else if carry = 2 then
35:      write 0
36:      have carry = 1 in mind
37: else if carry = 3 then
38:      write 1
39:      have carry = 1 in mind
40: if finished = 2 then
41:      You're done
42: else
43:      move to the far left
```

The following shows the state transitions for the main loop of this algorithm, the number of states would nearly double, if the $finished$ states would be taken into account.

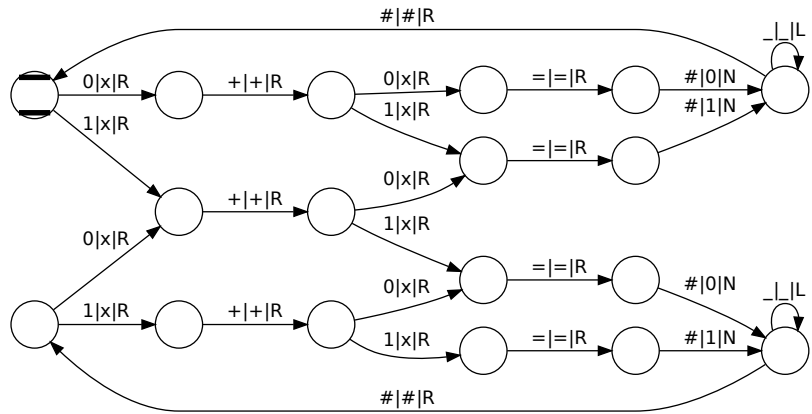

Seeing how complex turing machines get, it seems surprising, that it can code any machine or language.

### 2.2.2 WHILE is Turing complete

By definition 6, we would need a compiler from TM to WHILE, but by 3, an interpreter suffices. Implementing such an interpreter will be the following exercise.

#### ** Exercise 2    Interpreter for TM

1. Show that we can implement a dictionary datastructure in WHILE . Implement insert and lookup.
2. Show how you can code the transition function in your map.
3. Implement the tape
   a) Can you keep track of your position in a list with two stacks?
   b) Implement the left and right movement $[\![\text{left}]\!], [\![\text{right}]\!] : Tape \to Tape$. Note that the list is only potentially infinite, so you might in actuality run into the current edge.
4. Plug the pieces together to implement an interpreter turing(TM) for turing machines.
   a) How many nested while loops do you need?

#### Answer of exercise 2

1. Since execution time is not an issue, we can implement this as a list of pairs:
   ```
   lookup read (Table.Key) {
     for (K.Value) in Table {
       if [and](K = Key . [not](Result)) {
   ```

```
        Result := Value
      }
    }
  } Write Result

  insert read (Table.(Key.Value)) {
    Outtable := Nil
    for (K.V) in Table {
      if K = Key {
        Outtable := cons (Key.Value) Outtable
      }
      else {
        Outtable := cons (K.V) Outtable
      }
    }
  } write Outtable
```
   If you wanted to get rid of the copying of the whole list, you could also just prepend the new pair, since the lookup takes the first match.
2. The transition function $\delta : TapeAlphabet \times State \rightarrow TapeAlphabet \times State \times \{L, N, R\}$ can be coded by taking the alphabet and the states as symbols (or numbers) and storing `((in-letter.in-state).(out-letter.out-state.direction))` in the map M. Now $\delta = [\![spec]\!](lookup.M)$.
3. a) We can do that by keeping one stack `left` and one for `right`, the idea being, that the top of the `left` stack is the cell adjacent to the current to the left and the top of `right` is the same to the right. $Tape$ could now look like `(left.current.right)`
   b)
```
     left read (left.current.right) {
        right := cons current right
        if left {
          current := hd left
          left := tl left
        }
        else {
          current := :blank
        }
      } write (left.current.right)
```
   and analogously for `right`.
4.
```
   TM-step read (state.(left.current.right).transition-map)) {
     (state.current.dir) := [lookup](transition-map.(state.current))

     if (dir = :left) {
       (left.current.right) := [left](left.current.right)
     }
     if (dir = :right) {
       (left.current.right) := [right](left.current.right)
```

```
      }
    } write (state.(left.current.right))

    TM-run read (start-state.tape.transition-map.end-states) {
      current-state := start-state
      while [not]([contains](end-states.current-state)) {
        (current-state.tape) := [TM-step](current-state.tape.transition-map)
      }
    } write (current-state.tape)
```
This needs just the main loop.

### TM is WHILE-complete

It could now be, that WHILE is more powerful than TM, so that one can actually compute more with a WHILE program than one could with Turing Machines alone.

#### **** Exercise 3     Interpreter for WHILE in TM

In this exercise, we will build up a language that can be expressed in TM and will finally include WHILE. This is complex, but you might solve many low level implementation problems of programming languages in the progress.

1. Show an $n$-taped TM to the one taped (where $n$ is of course fixed at compilation time). A $n$-taped turing machine has $n$ tapes that can be independently moved and writen. *Hint: Unify the tapes and mark where you are.*

2. Show that for all TM's A and B exists a turing machine A;B such that $[\![A;B]\!]_{\text{TM}}(x) \simeq [\![B]\!]_{\text{TM}}([\![A]\!]_{\text{TM}}(x))$, i.e. that you can execute TM's one after another.

3. Show that you can implement a while construct
   a) Given a TM $A$ that prints T or F on a tape and another TM $B$, construct a TMif A {B} that runs $A$ and then $B$ only if the second tape shows $T$.
   b) Modify this so that $B$ gets executed as long as $A$ returns T.

4. Implement the cons datastructure on the tape of a turing machine. Note that you can add another tape to "take notes" by merit of the first question.
   a) Implement cons. *Hint: Be literal about expressions like* cons (cons nil nil) (cons nil nil).
   b) Implement hd and tl.

5. Show that you can implement a map, that can be used to save the variables.
   a) Given a name $a$ on one tape, can you look through a list of pairs and write $b$ on another tape if the coding of $(a.b)$ can be found in list.
   b) Implement changing the variables.

c) Argue, why it is not possible to simulate WHILE by using one tape per variable.

6. Argue, how you could now interpret WHILE.

<div style="text-align:center">**Answer of exercise 3**</div>

1. As we have only a limited amount of tape used, we can always move to the far left side of our tape. Now imagine a memory/tape layout like this (in the example for $n = 2$):

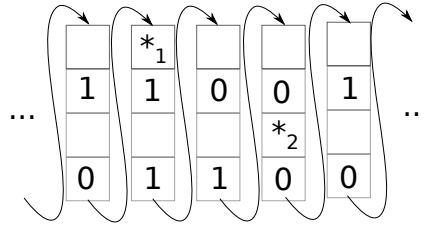

where the $*_n$ means *tape n is currently at the following value*.
Now design the states and transitions such that

1: **for all** tape $t$ **do**
2:     move to the far left.
3:     assume state $(s, t)$, where $s$ is the state of the multi-tape machine.
4:     move right until you find $*_t$.
5:     execute the step, that the multi-tape machine would make.
6:     move $*_t$ left or right, if necessary.

While this leads to a lot more states, it makes it possible to execute more than one tape simultanously.

2. A;B inherits the end-states of B and the starting-state of A. The internal states of A and B need to be disjunct, otherwise the machine could *jump over* from A to B or vice versa. To ensure that, we call the states in the new machine $(s, A)$ or $(s, B)$, depending on whether $s$ is from A or B. Now $\delta((s, A), x) = \delta_A(s, x)$ and $\delta((s, B), x) = \delta_B(s, x)$. Finally, we connect the two machines: $\forall e \in Endstates_A : \delta((e, A), x) = (start_B, x, N)$

3.   a) By the first question, we can assume, that A uses exactly one tape, and also that it is different from the one, that B uses. Use a similar construction as with the last question:

$$\forall e \in Endstates_A : \delta((e, A), (\_, T)) = (start_B, (\_, T), (N, N))$$

and

$$\forall e \in Endstates_A : \delta((e, A), (\_, F)) = (end, (\_, F), (N, N))$$

To round things up, connect the ends of B to the *end* state.
  b) Add the following constraint:

$$\forall e \in Endstates_B : \delta((e, B), \_) = (start_A, \_, (N, N))$$

4. a) We introduce the turingletters $\langle nil \rangle, \langle cons() , \langle\rangle\rangle, \langle,\rangle$ to the alphabet. Now given two tapes, we first write our return value on a third (empty) tape: first write $\langle cons()$ moving right, then we copy the first tape and add $\langle,\rangle$. Finally we copy the second tape and close with $\langle)\rangle$. Note here that cons and its friends are part of the WHILE language, and $\langle cons(), \ldots$ are letters we can put on a tape of a turing machine.

   b) Here we get a tape on which $\langle cons() \ldots \langle,\rangle \ldots \langle)\rangle$ is stored. The problem here is that we need to consider the nesting of cons. For this, we use an auxillary tape $aux$ and write a simple stack.

   1: write $x$ and move to the right on $aux$ if the current input is $\langle cons()$.
   2: move left on $aux$ if the current input is $\langle)\rangle$.
   3: start copying input to output after encountering ,.
   4: halt when the $aux$ tape reaches a blank after moving left.

   The machine for hd then is analogous.

   c) 1: **for all** p **do**airs $p$ on the tape

These two exercises show, that WHILE $\equiv$ TM. This is important, since when showing something to be WHILE-complete, we don't need to implement the WHILE, but can use TM instead, which is much simpler.

### 2.2.3 The Halting Problem

We saw in the introduction, that there are problems, that can be solved in WHILE, that can not be solved in FOR and also that serveral formalisms are equally powerful to WHILE, so that one can wonder, if this literally solves all problems.

Unfortunately, a simple cardinality argument shows that this is not the case: There are only countably infinitely many WHILE programs, as there are only so many finite strings, some of which are also WHILE programs.

The set of functions $\mathbb{N} \to \mathbb{N}$ however is bigger, as every freshmen course in mathematics will show, therefore, there must be some functions that can not be computed with WHILE programs. This proof is unsatisfying, as it does not give a problem that can't be solved. In fact, we would again describe problems in a finite way, so the argument is mood.

There are however many problems that can explicitely stated but not solved by any algorithm. The perhaps most notoric is the halting problem:

**Theorem 4** (Halting Problem). *There can be no program* halt? $\in$ WHILE *such that*

$$[\![\texttt{halt?}]\!] : \texttt{WHILE} \times I_{\texttt{WHILE}} \longrightarrow \{True, False\}$$

$$[\![\texttt{halt?}]\!](p, i) \simeq \begin{cases} False, & \text{if } [\![p]\!](i) = \bot \\ True, & \text{else} \end{cases}$$

So `halt?` *answers if a given program with a certain input will halt.*

*Proof.* The proof is a variation of the quintessential uncomputability proof: We assume that we had `halt?` and construct a program, that leads to a contradiction. The program is a variation of `inverse`:

```
loopInverse read P {
  if [halt?](P.P) {
    while True { }
  }
} write P
```

So we should get

$$[\![loopInverse]\!](p) = \begin{cases} \bot, & \text{if } [\![p]\!](p) \neq \bot \\ p, & \text{else} \end{cases}$$

what then is $[\![loopInverse]\!](loopInverse)$? Either it loops or it returns, so lets first assume that it loops:

$$[\![loopInverse]\!](loopInverse) = \bot \qquad \Rightarrow$$
$$[\![halt?]\!](loopInverse) = False \qquad \Rightarrow$$
$$[\![loopInverse]\!](loopInverse) = loopInverse \neq \bot$$

So on the other hand, what if it didn't loop?

$$[\![loopInverse]\!](loopInverse) \neq \bot \qquad \Rightarrow$$
$$[\![halt?]\!] = True \qquad \Rightarrow$$
$$[\![loopInverse]\!](loopInverse) = \bot$$

Therefore no such `WHILE` program can exist. □

Lets reflect on this proof: While it does use `WHILE`, it would be easy to translate `loopInverse` into other languages and give a halting problem for those. In fact, if we can compile `loopInverse` into a language, then the new language can not solve its own halting problem, even if it was more powerful than `WHILE`.

**Corollary 1.** *No language at least as powerful than* `WHILE` *can solve its own halting problem.*

### 2.2.4 Rice's Theorem

Now maybe the Halting Problem is very special and maybe we're not even that interested in the halting of the machine. Maybe we're interested in another property $P$, for example "if the machine halts, its result is the fibonacci frequence" or "the machine leaves this part of the memory untouched".

Such procedures would certainly be useful for specification, security, etc. but Rice proved in 1953 that these procedures cannot exist either.

**Theorem 5** (Rice). *If there is a property*

$$P : (I_{\texttt{WHILE}} \to O_{\texttt{WHILE}}) \to \{True, False\}$$

*such that there are procedures $x$ and $y \in \texttt{WHILE}$ such that $P(\llbracket x \rrbracket) = True$ and $P(\llbracket y \rrbracket) = False$ then there can be no procedure $\llbracket p \rrbracket : \texttt{WHILE} \to \{True, False\}$ that corresponds to $P$.*

*So if we have a non-trivial property of functions, we cannot write a program, that can deduce if the property holds based on the source of a procedure alone.*

*Proof.* We can easily see that the halting problem is a special case of this, but in fact, with a simple construction, all cases come down to it.

First we note that a procedure $\llbracket z \rrbracket (x) \simeq \bot$ must either have the property $P$ or not. Since $p$ halts, we can check that in advance. Assume that $\llbracket p \rrbracket (z) = False$, then we can build a program that can solve the halting problem. For this we use that there is a procedure $\texttt{known} - \texttt{p}$ that satisfies $P$ (by non-triviality). Then we can define

```
maybe-p: read (Q.I.X) {
  [interpreter](Q,I)
  Y := [known-p](X)
} write Y
```

so that $f = \llbracket spec \rrbracket (\texttt{maybe-p}, Q, I)$ satisfies the property $P$ if $\llbracket Q \rrbracket (I) \neq \bot$, otherwise $\llbracket f \rrbracket \simeq \llbracket z \rrbracket$.

```
halt: read (Q,I) {
  known-p-or-z := [spec](maybe-p.Q.I)
  Y := [p](known-p-or-z)
} write Y
```

so `maybe-p`

The trick here is that

$$\llbracket \llbracket spec \rrbracket (\texttt{maybe-p}, \texttt{Q}, \texttt{I}) \rrbracket (x) \simeq \begin{cases} \llbracket z \rrbracket, & \text{if } \llbracket Q \rrbracket (I) \simeq \bot \\ \llbracket known\text{-}p \rrbracket, & \text{else} \end{cases}$$

and therefore

$$\llbracket \texttt{halt} \rrbracket (q, i) \simeq True \Leftrightarrow \llbracket \llbracket spec \rrbracket (\texttt{maybe-p}, \texttt{q}, \texttt{i}) \rrbracket \simeq \llbracket \texttt{known-p} \rrbracket$$

$\Leftrightarrow \llbracket q \rrbracket (i)$ halts and vice versa for the contrary case. □

**So how do we deduce properties about programs then?**

Rice tells us, that it is not possible to deduce for example, that the result of a computation is a `String`. Yet the Java compiler does it all the time. How is this possible?

The trick is that Rice tells us that the problem can't be solved *in full generality*. So instead of accepting all programs, we accept only the ones for which we *can* prove the property. This throws away an infinitude of perfectly reasonable programs, which would keep the property, only because we can't prove it, but this way, we can be sure that for any program we accept, the property *does* hold.

### 2.2.5 The Normal Form Theorem

**Theorem 6** (Normal Form Theorem). *Every* `WHILE` *program can be transformed into a* `WHILE` *program with exactly one* `while` *loop. That is, all the inner constructs are taken from* `FOR`.

*Proof.* Lets call the language of `WHILE` programs, that are in the normal form $\text{WHILE}_1$. We have seen that we can transform $\text{WHILE} \to \text{TM}$ and $\text{TM} \to \text{WHILE}_1$. If we chain these (informal) compilation processes to get a compiler $\text{WHILE} \to \text{WHILE}_1$. □

### 2.2.6 Church's Thesis

As mentioned, when the notion of computability was first developed, there were competing formalisms, but as it turned out, they were all equally powerful. This lead Alonzo Church to conjecture, that what the human sees as intuitively computable, is adequately formalized by any of these competitors. This is known as Church's Thesis (or Church-Turing Thesis).

**Thesis 1** (Church's Thesis). *The Turing Machine adequately models what we hold for intuitively computable.*

We saw that `WHILE` too is a competitor, since it is equally powerful to any of the other formalisms. We could then formulate it as follows:

**Thesis 2.** *There is no intuitive extension of* `WHILE` *that is more powerful than* `WHILE`.

For example, we could add procedure calls, but that would not make any functions computable that were not computable before.

The thesis can not be formally proven, since it connects the informal concept of "intuitively computable" with the formal Turing Machines, however the evidence that no-one has since build a computer, that could be exploited to calculate more than a `TM` gives a strong indication, that the Church-Turing Thesis is true.

**. . . or is it?**

There has been some work, what would happen, if you added certain functionalities to Turing Machines (and by extension to `WHILE`).

Turing himself analysed so-called oracle machines, which are basically Turing Machines, but can answer specific questions to an oracle on a different tape, that would answer if a specific predicate is true for the value of the tape. The halting problem for Turing Machines is not solvable on a normal `TM`, but we could add an oracle for it. This would be more powerful than `TM`, but it is not intuitively clear how one could build such an oracle. Also, we would find, that the oracle machine could not solve its own halting problem – for very much the same reason that the original `TM` can't solve its own.

Another line of thought is that, when we talk about these formalisms, we don't really care if humans find it intuitively computable, but rather that we could actually build a machine that computes it. This gives us the physical Church-Turing thesis: It is not possible to build a computer that can not be simulated on a `TM`.

This is a thesis, that *can* be put to the test. While `TM` rely on classical mechanics, some thought has been put into quantum computing. The model that is currently used to build quantum computers (qubits) are however not stronger than `TM`. Faster yes, but not really stronger.

On a larger scale, it is possible to construct a computing device that orbits around a rotating black hole. An observer falling into the black hole could then observe the infinite computation time of the orbiting machine in subjectively finite time[7]. It has however been noted, that falling into a black hole might not be a good idea.

Finally, the notion of a computer running and halting after some time might be flawed. There is no procedure, that can compute $\pi$ to its full extend, but many approximation procedures, which can give a arbitrarily precise result. Is $\pi$ then intuitively computable?

### 2.2.7 Properties of Turing complete languages

There are some common elements in the languages that are Turing complete. As we have seen, any Turing complete language is unable to solve it's own halting problem – and by extension any halting problem of a Turing complete language.

Often, it is not obvious, that a certain formalism is Turing complete.

**Conway's Game of Life**

In 1970, the British mathematician John H. Conway devised a "game" on an infinite squared paper: we start with some squares marked as "alive" and then we update the paper in the following way[8]:

---

[7] See [3] for the whole discussion on the feasibility and the mathematical proof

[8] The general machine that updates cells only taking the cells neighbors into account is called a cellular automata

- A dead square becomes alive, if it has exactly three neighbors[9], that are alive. Imagine that the cells around it would reproduce and the offspring move into this square.

- If the cell is alive, it stays alive if only if it has two or three living neighbors.

These extremely simple rules still make up a turing complete language. How could one guess that it was?

The first hint is the discrete infinite data structure. We have an unlimited (but never infinite) number of squares that are alive.

This is a necessary condition to be Turing complete. Imagine, that we only allow a limited data structure, for example a Turing tape that has $n$ cells and an alphabet of $k$ letters. Each cell has therefore only $k$ possible configurations, and all the cells combined have only $k^n$ possible states. Since our machine only has a limited number of states, it can be in, say $m$, and the head has only $n$ posible positions to be in, after at most $n\,m\,k^n$ steps of the machine we must have reached a state that we have seen before. The transition function does only depend on these factors, therefore the loop will start again and nothing new will come of it.

Another hint is that there is no obvious way to deduce certain properties without running the whole machine – "Is this cell alive after $n$ turns" can only be anwered after running $n$ turns.

The real test is of course to implement a turing machine in the game of life, but this is beyond the scope of this text[10].

**Why are most programming languages Turing Complete?** Not being able to tell if a program will stop nor any other interesting properties seems like a bad start. Further, we use mostly algorithms that are guaranteed to hold in finite time. In short, we seldom truly need a Turing complete language, but still every major programming language is Turing complete.

The reason for this is that programmers don't want to think about the halting of their programs, because normally they obviously do. Proving that might not be trivial and might require the programmer to rewrite the program in such a way, that a computer can prove it. So while it would be possible to to implement most algorithms in a weaker language, it would be impractical. We couldn't use any unbounded loop constructs or true recursion.

In some areas, this is acceptable: In formal logic and interactive proof-finding, Coq[11] is used, even-though it has no unbounded recursion and every piece of code terminates.

---

[9]diagonal cells are neighbors too

[10]http://rendell-attic.org/gol/tm.htm has an implementation of a turing machine in the Game of Life

[11]http://coq.inria.fr/

## 2.3 Self Interpretation

Self interpretation is the ability of an formalism to support an "universal mechanism", that is a program that can interpret a finite description of any programs in itself and apply it to some input.

For computability, self interpretation can be seen as some kind of gold standard. This stems from the fact, that the simpler model of computation, FOR, are not self interpreting, which changes as soon as self interpretation is added.

### 2.3.1 Recursion Theorem

**Theorem 7** (Recursion Theorem). *For any procedure $p \in$ WHILE there is a procedure $p' \in$ WHILE so that $[\![p']\!](x) = [\![p]\!](p'.x)$. This can uniformly be computed by $Y \in$ WHILE, i.e. $[\![[\![Y]\!](p)]\!](x) = [\![p]\!]([\![Y]\!](p).x)$*

*Every procedure of WHILE might as well use its own source code.*

*Proof.* Instead of giving $Y$ and explaining how it works, here is how one could figure out how to do it[12]:

1. We start with $[\![h]\!](x) :\simeq [\![spec]\!](x.x)$, because that gives us

    $$[\![[\![h]\!](f)]\!](x) \simeq [\![[\![spec]\!](f.f)]\!](x)$$
    $$\simeq [\![f]\!](f.x)$$

    a procedure, that passes a given function as the first argument to itself. This is not the full solution though: It uses the original procedure $f$, not the transformed $[\![h]\!](f)$. Unfortunately, we can't just write $[\![h']\!](x) :\simeq [\![spec]\!](x.[\![h]\!](x))$, because that would give the same problem.

2. Instead we need to repeat this process

    $$[\![iterate\text{-}combinator]\!](combinator.program.x) :\simeq$$
    $$[\![program]\!]([\![spec]\!](combinator.(combinator.program)).x).$$

    What does this do? We run the program and pass as its first argument the code of a program that is modified by the combinator once. This is basically the step from above, only for *combinator*.

    $$[\![iterate\text{-}combinator]\!](c.program.x) \simeq [\![program]\!]([\![spec]\!](c.c.program).x)$$
    
    and to get the own source
    
    $$[\![[\![spec]\!](c.c.program)]\!](x) \stackrel{!}{\simeq} [\![iterate\text{-}combinator]\!](c.program.x)$$

---
[12]The proof here is based on [13]

3. From the equation above, we can see, that we need to wrap *iterate-combinator* onto itself, $[\![Y]\!]\,(program) :\simeq [\![spec]\!]\,(iterate\text{-}combinator.(iterate\text{-}combinator.program))$.

   This has the desired property:

$$\begin{aligned}[\![[\![Y]\!]\,(program)]\!]\,(x) &\simeq [\![[\![spec]\!]\,(iterate\text{-}combinator.(iterate\text{-}combinator.program))]\!]\,(x) \\ &\simeq [\![iterate\text{-}combinator]\!]\,(iterate\text{-}combinator.program.x) \\ &\simeq [\![p]\!]\,([\![spec]\!]\,(iterate\text{-}combinator.iterate\text{-}combinator.program).x) \\ &\simeq [\![p]\!]\,([\![Y]\!]\,(program).x)\end{aligned}$$

$\square$

**Example** (Quines). A *quine* is a program, that outputs its own source. It is a fun exercise for all students of computer science to find a quine in their favourite language. The existence of quines is ensured by the recursion theorem: Simply take the program $[\![id]\!]\,(x) \simeq x$ and then $[\![[\![Y]\!]\,(id)]\!]\,() \simeq [\![id]\!]\,([\![Y]\!]\,(id))$ is a quine.

This is unfortunately not how you program a quine. Few languages have a $Y$ procedure, and even if they had, they would use an own internal representation for the *program* and not the source code. To program a quine, you have to execute the steps of $Y$ by hand. Because `WHILE` does not handle strings, it might not be the best choice to implement the quine, so the following is written in `Python`, but the same principle could easily be applied to create a quine for other languages as well.

1. We basically need to return $[\![id]\!]\,(id)$ as a string, but in most languages, it's actually the other way around – we define a function and can decide to code it (as a string). That is commonly denoted $id(\ulcorner id \urcorner)$.

2. So let's define a function, that does the coding:

   ```
   def quote(source):
       return '"'*3 + source + '"'*3
   ```

3. Next, define a procedure, that prints its argument once without quotes, once with the quotes, so basically the literal string $id(\ulcorner \texttt{id} \urcorner)$.

   ```
   def quined(source):
       print source + '(' + quoted(source) + ')'
   ```

4. Finally, we can pass the source of the program to itself:

   ```
   def quote(source):
       return '"'*3+source+'"'*3
   def quined(source):
       print source + '(' + quote(source) +')'
   ```

```
    quined("""
    def quote(source):
      return '"'*3+source+'"'*3
    def quined(source):
      print source + '(' + quote(source) +')'

    quined""")
```

This recipe can be ported to many languages, including `C`, `Java` or `Haskell`.

**Why is it called the *recursion* theorem?** Since we have our own source, we can implement recursion with our interpreter:

```
fibonacci read (source.X) {
  if ([or](X = 0, X = 1)) {
    Y := 1
  } else {
    Y := [interpreter](source.(X-1)) + [interpreter](source.(X-2))
  }
} write Y
```

### 2.3.2 How this translates into logic

The notion of computability is closely related to that of decidability in logic. A statement is *decidable* if either itself or its negation can be proven. For many years, it was thought that given a set of axioms strong enough every statement in mathematics was decidable, but as we will see, that is not the case.

Without going too far into formal logic, I can only say that a successful proof of a statement can be seen as a trace of a program that finds the statement from the axioms. The other direction is true as well: A program can be understood as a description of a constructive proof,[13]. In short: Formal logic (over the natural numbers) is Turing complete!

Since a proof is nothing but a finite string of symbols, we can code it as a natural number. But since the domain, on which we use formal logic is natural numbers, this also means that we can produce predicates about other predicates, for example `the first argument is a proof of the second argument`.

This is where the recursion theorem comes in: According to it, and this is a bit informal, since first, we'd need to translate it into its logical equivalent, every predicate can assume, that its first argument is the coding of itself.

Then what happens, if we apply this to the predicate `"no number codes a proof of the first argument"`? We'd have that `"no number codes a proof of this statement"`. Assume that it is false, then there *is* a proof for the statement, but since the statement is false, we'd have a contradiction in our formal system. And if the statement is true? Then there is a true statement expressable in the system, that can not be proven. This is known as *Gödel's incompleteness theorem*.

---

[13] This equivalence goes extremely deep, see Curry-Howard correspondence, see [5] for a thorough introduction to proofs as programs.

# Chapter 3

# Complexity

As we saw, not all problems are solvable, but it seems not too far-fetched to say that a problem not being solvable by any algorithm is seldom a concern for most applications. Being solvable in a reasonable amount of time and space however can very easily become a problem. What use is a program that solves our problem, but takes hundreds of years to complete for any reasonable input? This is where the field of complexity theory sets in: it characterises the solvable problems by their "difficulty".

The most intuitive measure of difficulty is arguably time-complexity, i.e. how long any algorithm will take to solve the problem in the worst case. Also space-complexity needs to be considered, i.e. how much storage one needs to complete the computation.

Since bigger problems are harder to solve and bigger problems need more data to describe it fully, we analyse the asymptotic complexity in the size of the input, i.e. how the time and space needed develop as the input grows towards infinity. The most important distinctions here will be if there is a linear relation or a polynomial of higher order – or something worse, e.g. exponential time/space.

Another important classification is that of non-determinism: if we could guess during the computation, how would that affect the complexity of the problem? Could we solve things faster, if we only guessed good enough? Surprisingly the answer is that we are not sure (even though it is strongly suspected that being a good guesser really helps).

As in computability with the Church's thesis, there is the question of what is intuitively computable in a certain bound and even more than in computability, this question seems to rely on the chosen formalism, but most of the distinctions are quite robust. Still the question remains, if some extraordinary feat of engineering can build a machine that is much faster than our Turing machine model[1].

The problems we discuss are decision problems:

**Definition 10.** A problem $P$ is called a *decision problem*, if the required output is in $Bool = \{True, False\}$.

Our high-level `WHILE` language served us well in computability, but it would be more confusing to measure time and space requirements than on the lower-level Turing machine.

For example, how would we measure the used space in a simple, yet realistic way: If I set `Y := cons X X`, do I copy `X` twice? Once? Not at all? What happens if I reassign a variable, that was used earlier? This discussion would surpass the scope of this text[2].

**Definition 11** (Running time). For a machine $m \in $ `TM` and the input $x$, we will write $T_m^{\text{TM}}(x)$ or just $T_m(x)$ to denote the number of steps that $m$ needs to get from the starting state $q_0$ to the end state $e_{accept}$. If that does not happen, $T_m^{\text{TM}}(x) = \infty$.

---

[1]Quantum computing is faster than `TM`. A quantum computer can search a list for an element in $O(\sqrt{n})$ time as shown in [6].

[2] [8, p. 325f] gives a possible measure.

Not only time is a limited resource, also the memory that is used during the computation can limit the applications of an algorithm. Some problems could be solved for example just by reading the input and writing a fixed number of items into the memory – for example searching an element in a list –, others would need to increase their memory usage by a fixed amount each time the input doubles, for example calculating the median of a list of numbers, and some neen much more space.

**Definition 12** (space usage). The space usage $S_M(x)$ for $M \in \text{TM}$ is the maximum of the maximum of the space usage of the states of its computation or $\infty$ if the machine does not halt – even if it looped on the same cells over and over.

The space usage of a state of computation is the number of cells between the left-most and the right-most non-blank cell. For example `...# # 0 1 0 0 # 1 0 1 1 # # ...` would use 9 cells.

**Definition 13** (Complexity Classes).

$$\text{TIME}_f := \{l \in \text{TM} : \forall x \in I : T_l(x) \leq f(|x|)\}$$
$$\text{SPACE}_f := \{l \in \text{TM} : \forall x \in I : S_l(x) \leq f(|x|)\}$$

So

1. A program is in the complexity class $\text{TIME}_f$, if its running time is always bound by $f$ of the size of the input.

2. A program is in the complexity class $\text{SPACE}_f$, if its memory usage is always bound by $f$ of the size of the input.

And we call a problem $P$ in a complexity class $C$, if there is an algorithm $p$ that solves it, such that $p \in C$.

The most important complexity classes are those of the relatively slow growing polynomial functions:

**Definition 14** (Polynomial time and space).

$$\text{PTIME} := \bigcup_{p \text{ is polynomial}} \text{TIME}_p$$
$$= \left\{ l \in \text{TM} \exists p = \sum_{k=0}^{n} a_k \, x^k : \forall x \in I : T_l(x) \leq p(|x|) \right\}$$
$$\text{PSPACE} := \bigcup_{p \text{ is polynomial}} \text{SPACE}_p$$
$$= \left\{ l \in \text{TM} \exists p = \sum_{k=0}^{n} a_k \, x^k : \forall x \in I : S_l(x) \leq p(|x|) \right\}$$

That is:

1. A program is in the complexity class `PTIME`, if its running time is always bound by some polynomial.

2. A program is in the complexity class `PSPACE`, if its memory usage is always bound by some polynomial.

On the other hand the exponential functions grow so fast, that an algorithm taking exponentially long is often only feasible for small inputs.

**Definition 15** (Exponential time and space)**.**

$$\texttt{EXPTIME} := \bigcup_{p \text{ is polynomial}} \texttt{TIME}_{2^{p(x)}}$$

$$\texttt{EXPSPACE} := \bigcup_{p \text{ is polynomial}} \texttt{SPACE}_{2^{p(x)}}$$

So here

1. A program is in the complexity class `EXPTIME`, if its running time is always bound by an exponential of some polynomial.

2. A program is in the complexity class `EXPSPACE`, if its memory usage is always bound by an exponential of some polynomial.

## 3.1 The complexity hierarchy

We can clearly see that `LOGSPACE` $\subseteq$ `PSPACE` $\subseteq$ `EXPSPACE` just because of the functions involved, but the `TIME` and `SPACE` hierarchy is in fact interleaved, as the following theorems will prove.

**Theorem 8.**
$$\texttt{TIME}_f \subseteq \texttt{SPACE}_f$$

*Proof.* On our tape, the only way to increase the number of cells used is to write something in an blank cell. That we can do at most once per step. □

**Theorem 9.** *There are $a > 0$ and $q > 1$ and $\exp(x) := a \cdot q^x$ such that:*

$$\texttt{SPACE}_f \subseteq \texttt{TIME}_{\exp \circ f}$$

*Proof.*  1. Assume that $P \in \texttt{SPACE}_{f(x)}$, then $\forall x : S_P(x) \leq f(|x|)$.

2. Assume that $|x|$ is fixed, then in how many ways can the memory arranged to hold that property? Each cell can hold any of the symbols of $\Gamma$, so there are $|\Gamma|^{f(|x|)}$ ways to arrange that. Multiply that with the number of states that are not end states and you get the number of steps after either the machine is in a configuration it has seen before or goes to a new configuration – necessarily an end state. If the Turing machine is in the same configuration, then it will necessarily act in precisely the same way as before, get to the same configuration again and again, and therefore loop.

3. But since $S_P(x) \leq f(|x|)$, $S_P(x) \neq \infty$ and therefore

$$T_P(x) \leq (|Q| - 2) \cdot |\Gamma|^{f(|x|)}$$

□

**Corollary 2.**
$$\texttt{PTIME} \subseteq \texttt{PSPACE} \subseteq \texttt{EXPTIME} \subseteq \texttt{EXPSPACE}$$

Surprisingly, it is not known, if these inclusions are strict. It has not been proven yet, that $\texttt{PTIME} \neq \texttt{PSPACE}$ or that $\texttt{PSPACE} \neq \texttt{EXPTIME}$. Complexity theory is full of these uncertainties, even if it superficially mirrors computability theory, it has proven to be hard on much more basic levels than computability has.

### 3.1.1 Hierarchy Theorems

While not much is known, there are two important *hierarchy theorems*, which give proper inclusions. To formulate that, we need the notion of a *time constructible function*:

**Definition 16.** A function $f$ is called *time constructible* if there is a $M \in \texttt{TM}$, such that for sufficiently big $n$, $\texttt{TIME}_M(1^n) = f(n)$.

A function $f$ is called *space constructible* if there is a $M \in \texttt{TM}$, such that for sufficiently big $n$, $\texttt{SPACE}_M(1^n) = f(n)$.

Basically, this states, that there are Turing machines with this exact time or space requirement $f(x)$.

**Example.**   1. $f(x) = x$ is time and space constructible: Copy the input to the output tape.

2. If $f$ is space constructible, then $g(x) = (x+1) \cdot f(x)$ time and space constructible. Run the machine from above on the output of the constructing machine: Running the constructing machine takes $f(x)$ time and space and then running the linear machine adds $x \cdot f(x)$.

3. If $f$ is time/space constructible, then $\forall c \in \mathbb{N}^+ : g(x) = c \cdot f(x)$ is time/space constructible: Run the constructing machine $M$ for $f$, but cicle through $c$ states for each step of $M$.

4. If $f$ and $g$ are time/space constructible, then $h(x) = f(x) + g(x)$ is time/space constructible. Just run them one after the other.

5. $f(x) = \sum_{k=0}^{n} c_k \cdot x^k$ is space and time constructible for $c_i \in N^+$. Follows from above.

6. $f(x) = 2^x$ is time constructible: for the $k$th input symbol, count up to the number $1(0)^k$ in binary.

7. $f(x) = 2^x$ is space constructible: First write 1 on the output tape, then for each element of the input, copy the output tape to its own end.

8. If $f(x) \leq x$ for infinitely many $x$ then $f$ is *not* time constructible: We can't read the input, so we couldn't know, how long it would have been.

9. This is not true for space constructible functions: We could even read the input and not write anything, thus $f(x) = 0$ is time constructible.

**Theorem 10** (Time Hierarchy). *If $f$ is time-constructible, then*

$$\texttt{TIME}_{o(f(x))} \neq \texttt{TIME}_{f(x)^3}$$

*Proof.* 1. The proof will show, that we can find a function, that is in $\texttt{TIME}_{O(f(x)^3)}$, but not in any $\texttt{TIME}_{o(f(x))}$

2. The function that does that is

$$[\![H_f]\!](M, n) \simeq \begin{cases} True, & \text{if } \texttt{TIME}_M(n) \leq f(|(\mid n)) \text{ and } M \text{ ends in } q_{accept} \\ False, & \text{otherwise} \end{cases}$$

We can run $H_f$ by using an interpreter ($O(steps^2)$) and counting the steps on a separate tape, so pessimistically[3] this is in $\texttt{TIME}_{O(f(x)^3)}$.

3. On the other hand assume that $H_f \in \texttt{TIME}_{f(\lfloor \frac{|x|}{2} \rfloor)}$, then $[\![G]\!](m) :\simeq$ not $[\![H_f]\!](m,m)$ runs only

$$f\left(\left\lfloor \frac{2|m|+1}{2} \right\rfloor\right) = f(|m|)$$

steps.

4. Thus
$$[\![G]\!](G) \simeq not\ [\![H_f]\!](G,G) \simeq not\ True \simeq False,$$

but that would imply, that $H_f$ would reject it, therefore

$$[\![G]\!](G) \simeq not\ [\![H_f]\!](G,G) \simeq not\ False \simeq True$$

and vice versa – which is a contradiction. Therefore $G \notin \texttt{TIME}f(x)$ and $H_f \notin \texttt{TIME}f\left(\lfloor \frac{x}{2} \rfloor\right)$ □

**Corollary 3.**
$$\texttt{PTIME} \neq \texttt{EXPTIME}$$

*Proof.* $x^n$ is time constructable for all $n \in \mathbb{N}$, but $x^{3n} < 2^x \Leftrightarrow 3n \cdot \log_2 x < x \Leftrightarrow \frac{\log_2 x}{x} < 3n$, which holds for any $n$, if $x$ is big enough, so there is no polynomial close enough to the exponentials, that would allow the $\texttt{PTIME}$ class to "jump" into the $\texttt{EXPTIME}$ class. □

---

[3]The Time Hierarchy Theorem holds even for $\texttt{TIME}_{f(x) \cdot \log f(x)}$ with a trickier interpreter and counting

**Theorem 11** (Space Hierarchy). *If $f$ is space constructible, then $\texttt{SPACE}_{o(f(x))} \neq \texttt{SPACE}_{f(x)}$.*

*Proof.* This proof is very similar to the Time Hierarchy theorem. Instead of counting the steps, we can write out $f(x)$ in unary on a second tape, and move this in parallel to the computation tape. If at any point, an unmarked cell is hit on the second tape, we know that we used more than $f(x)$ cells and reject. □

This theorem is much stronger than the time hierarchy theorem, because it shows that any difference in the asymptotic behaviour leads to a different computability class.

## 3.2 Nondeterminism

Up until now, choice did not come into play at any moment: A program gave a unique trace of executed instructions for any given input. In many cases however, algorithms contain a moment of arbitrariness, where in the deterministic case, one fixed way would need to be chosen. The run-time of the algorithm can depend very much on the choices we make.

**Definition 17.** A non-deterministic Turing machine (`NTM`) is a deterministic Turing machine, where the transition function $\delta$ returns sets instead of values now.

$$\delta : Q \times \Gamma \to \mathcal{P}(Q \times \Gamma \times \{L, N, R\})$$

The semantics need to adapt to that, a configuration of a non-deterministic Turing machine needs to contain serveral configuration of a deterministic run:

**Definition 18.** A *configuration* is $C \subset \Gamma^* \times \Gamma \times \Gamma^* \times Q$ and a step by a `NTM` is $C_1 \vdash C_2$ if
$$C_2 = \bigcup_{c \in C_1} \{step(tape_c, s_c, t) : t \in \delta(c)\}$$

This allows us to define

$$[\![M]\!]_{\texttt{NTM}}(tape_1) = \bigcup_{(tape_1, q_0) \vdash^*_M C} \{tape : (tape, q_{accept}) \in C\}$$

So if any computation for some choices ends in $q_{accept}$, then the computation of the non-deterministic machine succeeds. For decision problems, we only need to know that this set is not empty.

**Example.** When parsing the string `aabab` with the regex (a|ab)+, we can't know after the first letter in which branch of the *or* we will be. Only after we saw, that a second `a` follows, we know that it was the first one.

**Definition 19.** The time measure `NTIME` is the *minimum* of the `TIME` measures of the possible traces. Analogously, the space measure `NSPACE` is the minimum of possible `SPACE` measures of the traces. It denotes the fastest/least space consuming possible way to solve the problem, if guessing is allowed.

At first, this seems like a new mode of computation and thus belong to computability rather than complexity, but just proving that *some* computation accepts the input can be done by performing a breadth-first search on the possible traces. The complexity however seems very different, because in the worst case, this approach will take $\mathcal{O}(2^{\mathcal{O}(n)})$ deterministic steps per non-deterministic step, thus an algorithm that takes `NTIME`$f(x)$ will take `TIME`$2^{\mathcal{O}(f(x))}$ with this approach.

The most important class, that uses non-determinism is `NPTIME`: The problems, that can be solved in polynomial time, if guessing is allowed and we guess optimally.

**Theorem 12** (Guess and check)**.** *If $X \in$ `NPTIME`, then we can find a so-called certificate $w \in \Gamma^*$ for each input $x$, with $|w| \leq p(|x|)$ for a polynomial $p$, so that there is a deterministic Turing machine $M$ that decides $X$ given $w$ and $x$ in `PTIME`. If we can formulate the problem in such a way, then it is in `NPTIME`.*

*This means, that we can check our hypotheses for positive answers in polynomial time, so the hard part is only coming up with the good hypotheses. There is however no requirement for negative solutions, so the algorithm might even fail by looping forever.*[4]

*Proof.* On one hand, we can code the trace, that lead to the succeeding computation, to the tape. Checking, that a trace is valid for a given Turing machine is trivial.

If on the other hand we can produce a certificate for any input and check it in polynomial time in $|x|$ alone, then that certificate can only be $p(|x|)$ long for some polynomial $p$ – otherwise, we wouldn't be able to check the certificate this fast.

But then, we can define a non-deterministic machine, that comes up with this input in $p(|x|)$ steps and checks that afterwards by simulating the deterministic checker. □

We can see that `PTIME` $\subset$ `NPTIME`, by just outputting singleton sets in the non-deterministic $\delta$ and intuitively it seems that non-determinism must accelerate the computation significantly.

**Thesis 3.** `PTIME` $\neq$ `NPTIME`

However, up to this point, every try to prove or disprove this statement failed.

---

[4]The class of decision problems, that fail in non-deterministic polynomial time, but might run forever to check for success is called $co -$ `NPTIME`.

## 3.3 Complexity in Languages

While most languages we encounter are turing-complete, that does not need to be the case. In this chapter, we will discuss the notion of a problem description as a programming language.

**Problem descriptions**

There are many different kinds of problems in computer science, for example sorting a list or determining if a boolean formula with variables is always true. Nevertheless, these kind of problems are never singular – we would not want an algorithm[5], that could reliably sort the list $4, 8, 2, -9$, but one that could sort *all* lists, no matter the size. This means of course, that algorithms need data as input to describe the problems they need to solve. This data is then called the *problem description*.

Looking back at the beginning, the semantic function was introduced to give meaning to data. Since then, we primarily used it to differentiate between `WHILE` programs and the functions they denote. In the case of problem descriptions, we can the interpretation of a problem would be the solution that we would expect.

**Example.** $U = Cons(Nil, Cons(Nil, Cons(Nil, Nil)))$, then $[\![U]\!]_{unary} = 3$.

**Example.** $L = Cons(4, Cons(8, Cons(2, Cons(-9, Nil))))$ is the problem description for sorting the list $4, 8, 2, -9$ if given to a sorting procedure, formally $[\![L]\!]_{sort} = Cons(-9, Cons(2, Cons(4, Cons(8, Nil))))$. When given to a procedure, that calculates the minimum, the interpretation would be that instead, so $[\![L]\!]_{min} = -9$.

In this light, a problem description becomes a small and domain specific programming language, with the algorithm that solves it being an interpreter.

### 3.3.1 Reductions

Often in courses on algorithms, the same algorithm can be used in many different domains, because it has been observed that the two problems are *essentially the same* or that one is *essentially a special case of another*.

For example in chapter 2.2.4, we saw that we could formulate any non-trivial function property as a halting problem.

**Definition 20.** For a complexity class $X$, a *$X$-reduction* of a problem $A$ to a problem $B$ is a compiler from the problem descriptions of $A$ to the descriptions of $B$, so that compiling a description of $A$ and executing it as a $B$ problem description is in $X$. We write $A \leq_X B$.

**Example.** The most common reductions are:

---

[5]Most definitions would not even allow this as an algorithm

1. `PTIME`-reductions translate and execute a problem in polynomial time and therefore `NPTIME` and `PSPACE` are closed under these reductions. Nearly every `PTIME` problem can be `PTIME`-reduced to any other[6].

2. Linear time reductions give much stronger bounds, so that for example a $O(n^2)$ problem to which another is linearly reduced, proves that the other problem is $O(n^2)$ as well.

3. Computable reductions are used to show the computability or incomputability of problems.

### 3.3.2 Examples

The following examples should give a good impression of how the results of complexity can be used and what typical problems can look like. Since diversity was important in the writing, some results use techniques and helpers that have not been introduced here, but should be familiar for many nonetheless.

**Problems**

**Factorization of binary numbers** Given an integer $N$ to be factored and an upper bound $1 < M < N$, is there a $1 < d \leq M$ so that $\exists x \in \mathbb{N} : N = d\,x$. Or put in another way: is there a non-trivial divisor for $N$, that is no bigger than $M$.

It is obvious that the problem is contained in `NPTIME` by the guess-and-check method. At first it might seem, that it is actually linear time: Just check up to $M$ if the counter divides $N$, but since $M$ is written in binary, it only takes $\log_2(M)$ space for the input, therefore counting up to the number is actually exponential in the input.

**Example** (Relevance)**.** The RSA algorithm for public key encryption allows to deduce the private keys from the public keys – but only, if the attacker can factorize a big integer.

**Context Free Grammar membership (CFG)** A context-free grammar is a set of rules, how non-terminal symbols can be replaced by a mixture of again non-terminal symbols or terminal symbols (which can not be changed afterwards).

An easy example is the language of balanced braces:

$\langle word \rangle ::= \langle empty \rangle$
$\quad | \quad \text{`('} \langle word \rangle \text{`)'} \langle word \rangle$
$\quad | \quad \text{`['} \langle word \rangle \text{`]'} \langle word \rangle$
$\quad | \quad \text{`\{'} \langle word \rangle \text{`\}'} \langle word \rangle$

---

[6] All instances that have accepting and rejecting answers

which contains words like ([()()]{}) but not (().

The problem is now, given a set of rules $R$, a start-token $A$ and a string $s$, to check whether $s$ could be produced from $A$ under the rules $R$. Without loss of generality, we can assume that each step produces at least one non-terminal character*(citation needed)* .

This is surely in `NPTIME`, because if we were given the steps that expand $A$ to $s$, we could check that they do indeed produce $s$. There are however serveral algorithms, that efficiently expand the currently possible interpretations and can find the solution in $\mathcal{O}\left(|s|^3 \, |R|\right)$ time[7] and therefore the problem is also in `PTIME`.

**Example** (Relevance)**.** The syntax of programming languages is often described as a context free grammar (for example the syntax of `FOR` in table 1.1).

**Traveling Salesman Problem (TSP)** A salesperson wants to travel to a number of locations across the country, but is interested in getting home as soon as possible. Given a map with travel times between the cities, is it possible to do the trip in time for their wedding anniversary in $k$ days?

More abstractly: Given a graph with weighted edges, is it possible to find a path crossing all nodes, so that the sum of the weights of the edges crossed is no bigger than $k$?

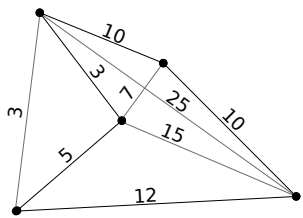

**Example** (Relevance)**.** A circuit is basically a round-trip over a board. As such, minimizing the resistance, ... encompasses solving the TSP.

**Satisfiability of boolean expressions (SAT)** Given a boolean expression with variables, is there a way to assign the variables, so that the whole expression is true?

**Example** (Relevance)**.** When modelling certain domains, for example in a system of artificial intelligence, constraints are often expressed as boolean expressions or can be converted to them[8]. Checking whether the constraints can be satisfied at all can already solve problems: If we have the facts $F$ and the hypothesis $H$, then $H$ fits the facts, if and only if $F \wedge \neg H$ is *not* satisfiable.

---

[7]See [12]
[8]See [11]

**Polynomial reductions**

SAT is an extremely handy problem to reduce to, so the following reductions will mostly feature that.

**SAT reduces to TSP**   It can be shown, that SAT can be reduced to a special case of itself, namely $3-cnfSAT$, the satisfiability of formulas in the so-called 3-clause normal form

$$\bigwedge_{i=1}^{n} a_i^1 \vee a_i^2 \vee a_i^3$$

. where $a_i^j$ can be any $x_k$ or $\neg x_k$. The $a_i^1 \vee a_i^2 \vee a_i^3$ are known as the *clauses*.

How this could be done might become apparent when looking at the proof of theorem 14. Without loss of generality then, we can assume, that the formula is in this form, when we try to reduce it to TSP. This has the advantage, that we know, that all clauses must be true and that in all clauses, one of the atoms must be true.

TSP takes as an input the maximum summed cost, but it is clear, that there is a special case, in which only the existance of a round-trip is asked for, not the limit of the cost[9]. This case is known as the Hamiltonian Cycle problem.

Now we need to create a graph, that is polynomially large in the length of the formula, but contains a round-trip if and only if the formula is satisfiable. The trick is of course using the clause-normal-form: Represent all clauses as a node $n_k$. The idea is now to only allow connections to $n_k$, if the representation of the atom makes the clause true[10].

Variables have a more complex structure

- a start node, in which the decision takes place, if the variable is set to $True$ (left) or $False$ (right). Depending on choice, the steps through the middle will allow of disallow checking off clauses.

- a bridge in the middle, that can either be just walked through, but also has some connections to the $n_k$ nodes that will be explained shortly.

- an end node, which will be the start node to the next variable.

It is clear, that the only round-trip possible leads through assigning all variables true or false, but we don't have a coding for *variable i appears in clause j*.

This is done with giving meaning to the hops on the bridge: If the variable $i$ makes $j$ true if taken as $True$, then the $j$th such pair contains a hop from the left node to the right one over $n_j$, and from the right to the left, if assigning $False$ makes $j$ true.

---

[9]One could for example set all costs to zero and check for any number $M$

[10]This construction is based on [12, p. 286-291]

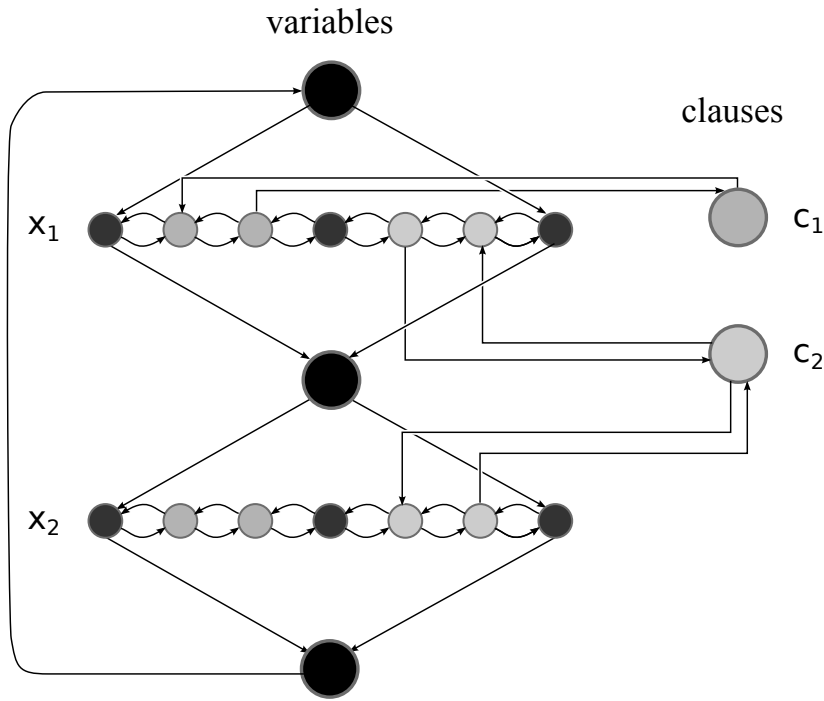

Figure 3.1: The tour for the expression $(x_1 \vee \neg x_2) \wedge x_2$

**Theorem 13.** *This construction has a round-trip if and only if the original formula is satisfiable.*

*Proof.* If the formula is satisfiable, then we can tick off one clause after another, by taking the hops from the bridge to $n_k$ whenever possible.

Now if the formula is *not* satisfiable, then that means, that at least one clause cannot be made true, but that translates directly to not bein able to jump over to it with the current variable assignment in the graph. □

**TSP reduces to SAT**  Lets reduce the TSP for the graph $G = (Nodes, Edges)$ for at most the cost $M$ to a problem of satisfiability of boolean expressions.

We first notice, that we can add and compare numbers by building a fixed length binary calculator from the logical connectives, that is able to add all weights if necessary. This won't require to introduce more than polynomial many variables and connectives.

Let the variables $p_i^n$ denote, that the $i$th node of the path is $n$, then

$$PathExclusion = \bigwedge_{i=1}^{|Nodes|} \bigwedge_{n \in Nodes} \bigwedge_{\substack{n' \in Nodes \\ n' \neq n}} p_i^n \rightarrow \neg p_i^{n'}$$

says that at any given position, only one node can be visited. Now we can just add up all weights multiplied with 1 if they are in the round-trip or 0 if they are not. This would look messy, because we have to do it bitwise, but the solution would look like

$$PathExclusion \wedge \left( \sum_{e=\in Edges} cost(e) \cdot include\text{-}edge(e) \right)$$

where

$$include\text{-}edge(n_0, n_1) = \begin{cases} 1, & \text{if } \bigvee_{k=1}^{|Edges|} p_k^{n_0} \wedge p_{k+1}^{n_1} \\ 0, & \text{otherwise} \end{cases}$$

This formula would still be polynomially long in the input length and wouldn't contain any predicates. Therefore, it is polynomially reduced to $SAT$.

**Factorization reduces to TSP** Factorization would be very easy to reduce to SAT: Given a $n$ digit binary number $z$, we could device $a_i$ and $b_i$ for $i = 1 \ldots (n-1)$ as $n-1$ digit binary numbers. We could then build a binary multiplier[11] and check, if the result is the same as $z$. But we already found, that any SAT problem can be reduced to TSP, so if we use the construction with the zick-zack-gates again and on a large scale, we can convert $a_i$ and $b_i$ and the binary multiplier. Since applying two polynomial reductions in series still is a polynomial reduction, we can reduce Factorization to TSP.

**CFG reduces to SAT** Here as well, it is useful to use a special form, in order to manage the complexity that the problem in unrestricted form offers: Context-free grammars can (relatively easily with a mere $n^4$ blow up of the grammar) be converted to the so called Greibach Normal Form, in which rules always have one of the following forms:

$$X \rightarrow \alpha, A_1, A_2, \ldots, A_n$$
$$S \rightarrow \varepsilon$$

where $A_i$ are non-terminals and $\alpha$ is a terminal. Furthermore $\varepsilon$ denotes the empty string and $S$ the start symbol.

This form has some consequences, for example that only one rule if any can produce empty strings, all others produce at least one terminal. Another nice

---

[11] Consult your favourite book on computer engineering

property is that the string grows to the right, we don't *need* to go back if we perfer not to.

In this case, we will just expand the left-most non-terminal one after another until the string is fully produced – after at most $n$ steps, since we don't need to use the empty string rule, except when explicitly matching the empty string. Now let $R_t^r$ denote that we use the rule $r$ at the step $t$. Since we expand one after another, we need to formalize exclusion as seen earlier:

$$RuleExclusion = \bigwedge_{t=1}^{n} \bigwedge_{r \in Rules} \bigwedge_{\substack{r' \in Rules \\ r' \neq r}} R_t^r \to \neg R_t^{r'}$$

Furthermore we need to model the string after the step $t$, by defining $S_{x,t}^A$ to mean that the $x$th symbol of the string is $A$ at the time $t$. Again, we exclude the possibility, that the string has multiple symbols at the same time and space.

$$SymbolExclusion = \bigwedge_{t=1}^{n} \bigwedge_{x=1}^{n} \bigwedge_{s \in Symbols} \bigwedge_{\substack{s' \in Symbols \\ s' \neq s}} S_{x,t}^s \to \neg R_{x,t}^{s'}$$

With these at our disposal, we can define the expansion rules: $Rule_j = X \to \alpha, A_1, \ldots, A_m$ becomes $\ulcorner Rule_j \urcorner = S_{i,t}^X \to S_{i,t+1}^\alpha \wedge \bigwedge_{k=1}^{m} S_{i+k,t+1}^{A_k}$

Then the matching of context-free grammars can be solved by solving

$$RuleExclusion \wedge SymbolExclusion \wedge \bigwedge_{t=1}^{n} \bigvee_{j=0}^{|Rules|} \ulcorner Rule_j \urcorner$$

### 3.3.3 Hardness and completeness

We saw that in computability, there seems to be a natural border of computability, in which `WHILE`, Turing machines and nearly any reasonably strong language reside. It was not too surprising then, that these different languages could express each other and themselves by means of interpretation and compilation. Maybe more surprising is, that the complexity classes `PTIME` and `NPTIME` seem to be a similarly natural border.

**Definition 21.** For a complexity class $C$, a problem $P$ is *hard*, if $\forall c \in C : c \leq_{\texttt{PTIME}} P$.

The problem $P$ is $C$-*complete*, if $P$ is $C$-hard and $P \in C$.

**Example.**
- `FOR` is `PTIME`-hard, but not complete, because not all `FOR` programms run in `PTIME`.
- `TM` is `WHILE`-complete.

These examples are a bit cheated, since they use extremely large complexity classes. But in fact, we have seen two problems already, that are `NPTIME`-complete.

**SAT is `NPTIME`-complete**

**Theorem 14.** *The satisfiability of boolean formulas is `NPTIME`-complete.*

*Proof.* For completeness, inclusion and universality are required.

To prove $SAT \in \texttt{NPTIME}$, we use the guess-and-check definition of `NPTIME`. The required certificate in this case is a mapping of the variables to truth-values. This is at most linear in the length of the expression, therefore fulfilling the required `PSPACE`length of the certificate. Given the certificate, we can then fully evaluate the expression, which again takes linear time.

The general idea how to create an $X$-complete problem is to simulate the mode of computation. In the case of space completeness, you fail if you move out of a premarked part of the tape, in the case of `PTIME`-completeness, you count down your remaining time and in the case of `NPTIME`-completeness, you build a table of possible configurations of the turing machine and see if any of them succeeds[12].

First of all, it is not necessary to use transition function $\delta : Q \times \Gamma \to \mathcal{P}(Q \times \Gamma \times \{L, N, R\})$, it is sufficient to be able to choose between two alternatives. If for example $\delta(q, x) = \{(q_1, y_1, d_1), \ldots, (q_n, y_n, d_n)\}$, then we can simply and polynomially reduce it to a list of binary choices[13]:

$$\delta(q, x) = \Big((q_1, y_1, d_1), (q^{(1)}, x, N)\Big) \tag{3.1}$$

$$\delta(q^{(1)}, x) = \Big((q_2, y_2, d_2), (q^{(2)}, x, N)\Big) \tag{3.2}$$

$$\vdots \tag{3.3}$$

$$\delta(q^{(n-1)}, x) = \Big((q_{n-1}, y_{n-1}, d_{n-1}), (q_n, y_n, d_n)\Big) \tag{3.4}$$

So without loss of generality, we can assume, that the non-deterministic Turing machine makes at most binary choices, which we could code as booleans. Further, if we *don't* want to make a choice, we could make a choice, but end in $q_{reject}$ as the second option, so we can make *exactly* one binary choice per step.

So let us now reduce the arbitrary problem $A \in \texttt{NPTIME}$ to $SAT$. Since it is polynomial runtime, there is some $k$ for which the runtime is bound by $p(n) = n^k$.

Even assuming, that we made a choice in every step, it would be enough to give $p(n)$ booleans $c_1, \ldots, c_{p(n)}$ to fully specify the trace. This will be, what the solver for SAT would then modify to find the solution to $A$, the rest is deterministic.

The coding, if a trace ends in an accepting state, remains to be done – which includes coding how a Turing machine works.

The state changes over time, and has $|Q|$ exclusive possibilities. Then $s_t^q$ means, that the machine is in state $q$ at the time $t$. Coding, that the possibilities

---

[12]This proof combines ideas from the proofs given by [8] and [12]

[13]The use of tuples instead of sets stems from the fact, that one cannot speak of the *first* or *second* element of a set, but for strict semantics, this is necessary.

exclude each other can be done by adding

$$StateExclude = \bigwedge_{\substack{q \in Q \\ t=1\ldots p(n)}} \bigwedge_{\substack{q' \in Q \\ q' \neq q}} s_t^q \to \neg s_t^{q'} \qquad (3.5)$$

$$= \bigwedge_{\substack{q \in Q \\ t=1\ldots p(n)}} \bigwedge_{\substack{q' \in Q \\ q' \neq q}} \neg s_t^q \vee \neg s_t^{q'} \qquad (3.6)$$

This gives $|Q|^2 \, p(n)$ additional input characters – nothing that would break polynomial input to $SAT$.

The tape only needs $p(n)$ cells, but for each of the $p(n)$ time steps, therefore we code the tape as the table $T_{x,t}^c$ with $c \in \Gamma$, $x = 1\ldots p(n)$ and $t = 1\ldots p(n)$, meaning *the tape cell x contains the character c at the time t*. The constraint is of course, that at any given time, the tape at a given cell can only contain one character, but this can be expressed by

$$TapeExclude = \bigwedge_{g \in \Gamma} \bigwedge_{\substack{x=1\ldots p(n) \\ t=1\ldots p(n)}} \bigwedge_{\substack{g' \in \Gamma \\ g' \neq g}} T_{x,t}^g \to \neg T_{x,t}^{g'} \qquad (3.7)$$

$$= \bigwedge_{g \in \Gamma} \bigwedge_{\substack{x=1\ldots p(n) \\ t=1\ldots p(n)}} \bigwedge_{\substack{g' \in \Gamma \\ g' \neq g}} \neg T_{x,t}^g \vee \neg T_{x,t}^{g'} \qquad (3.8)$$

. Despite looking huge at first glance, this is still polynomially long in the input.

We also need to keep track of where the read/write head of the Turing machine is. Perhaps unsurprisingly, we introduce a table $H_{x,t}$, meaning that at the time $t = 1\ldots p(n)$ the head is at the position $x = 1\ldots p(n)$. Exclusion applies, so we add

$$HeadExclude = \bigwedge_{\substack{x=1\ldots p(n) \\ t=1\ldots p(n)}} H_{x,t} \to \neg H_{x,t} \qquad (3.9)$$

$$= \bigwedge_{\substack{x=1\ldots p(n) \\ t=1\ldots p(n)}} \neg H_{x,t} \vee \neg H_{x,t} \qquad (3.10)$$

.

Finally we need to code how the transitions take place. We need to say that

- Deterministically changing state means, that if we were in state $g$ to time $t$ and saw the cell $T_x$ with the character $c$, then in $t+1$, we are in the state $g'$ implied by $\delta$.

$$StateTransition^{q'}(s_t^q, T_{x,t}^c) = s_t^q \wedge T_{x,t}^c \to s_{t+1}^{q'} \qquad (3.11)$$

$$= \neg s_t^q \vee \neg T_{x,t}^c \vee s_{t+1}^{q'} \qquad (3.12)$$

- Deterministically changing the current cell does the same thing, only with cells instead of states.

$$CellTransition^{c'}(s_t^q, T_{x,t}^c) = s_t^q \wedge T_{x,t}^c \to T_{x,t+1}^{c'} \qquad (3.13)$$

$$= \neg s_t^q \vee \neg T_{x,t}^c \vee T_{x,t+1}^{c'} \qquad (3.14)$$

- Deterministically moving on the tape only changes $H_{t+1}$

$$MoveTransition^d(s_t^q, T_{x,t}^c) = s_t^q \wedge T_{x,t}^c \wedge H_{x,t} \to H_{x+d,t+1} \quad (3.15)$$
$$= \neg s_t^q \vee \neg T_{x,t}^c \vee H_{x,t} \vee H_{x+d,t+1} \quad (3.16)$$

, where $d \in \{-1, 0, 1\}$

- If our current choice bit says $True$, then take the first option $X_t$, else the second option $Y_t$. This could for example be expressed by $(c_t \wedge X_t) \vee (\neg c_t \wedge Y_t)$.

- Plugging this together, we get the coding of the transition function.

$$FirstTransition_{x,t} = H_{x,t} \wedge T_{x,t}^c \wedge s_t^q \to$$
$$MoveTransition^d(s_t^q, T_{x,t}^c) \wedge$$
$$CellTransition^{c'}(s_t^q, T_{x,t}^c) \wedge$$
$$StateTransition^{q'}(s_t^q, T_{x,t}^c)$$
$$(3.17)$$

if $\delta(c, q) = ((c', q', d), (c'', q'', d'))$ and analogously for the second choice

$$Transition = \bigwedge_{t=1}^{p(n)} \bigwedge_{x=1}^{p(n)} (c_t \wedge FirstTransition_{x,t}) \vee (\neg c_t \wedge SecondTransition_{x,t})$$
$$(3.18)$$

Even this huge transition table is still polynomial in the length of the input.

Finally, the formula should only be accepted of the end state is $q_{accept}$, which is simply $EndState = s_{p(x)}^{q_{accept}}$.

Now $Transition \wedge HeadExclude \wedge TapeExclude \wedge StateExclude \wedge EndState$ is the representation of $A$ in $SAT$. Note that no functions (predicates) would have been necessary[14].

With this so-called *gadget* – something that simulates the computational behaviour of another problem – we have reduced the problem $A$ to $SAT$. Basically we reduced the problem of calculating $p(n)$ steps on a non-deterministic Turing machine, which is by definition `NPTIME`-complete, to the $SAT$ problem. □

This implies, that SAT can express any other `NPTIME` problem with only a polynomial slowdown. This also means, that SAT could model any `PTIME` problem, such as checking context free grammar membership.

On the other hand, we saw that SAT can be reduced to the Traveling Salesman Problem, which shows its completeness as well.

---

[14]Slightly changing this proof shows then, that predicate logic is actually Turing complete.

There are indeed many relevant problems that are NPTIME-complete[15], for example maximizing linear functions over linear constraints in the domain of integers or optimal scheduling of tasks over multiple processors just to name two.

#### NPTIME and optimization

While one should keep in mind, that NPTIME only describes decision problems, many NPTIME-complete problems actually hide an optimization problem. For example, if we can determine if the traveling salesman can do the round-trip with cost of at most $k$, then we can actually find the optimal path:

1: find least $k_0$, so that the round-trip is possible with binary search over $k$
2: **for** each edge $e$ in the graph **do**
3:     Remove $e$ from the graph
4:     Check if the round-trip cost increases
5:     If so, reinsert $e$

After the algorithm finishes, the only edges in the graph are the ones in the round-trip.

Why does this work? First the binary search determines the exact cost of the optimal route and does so in only $\log k_0 \cdot f(x)$, where $f(x)$ is the time that TSP takes.

The loop then uses that any edge, that is not in the round-trip will leave the cost untouched, even when removed. An edge, that can not be compensated however will increase the cost, when removed. If there are more than one possible round-trip with the same cost, then only one of them will survive the loop, because edges will be removed, if any of the routes is still possible. We would need $|E| \cdot f(x)$ time for the loop.

All in all this algorithm will take $(\log k_0 + |E|) \cdot f(x)$ and therefore has only a linear slowdown compared to solving TSP.

#### Help, my problem is NPTIME-complete

While being (probably) asymptotically very difficult to solve, NPTIME-complete problems are by no means impossible. There are some strategies that can be applied to handle them.

**NPTIME-complete does not necessarily mean slow**  In many cases, the precise algorithm to solve the problem works reasonably fast in most cases, but explodes for some abnormal cases. If these abnormal cases don't occur in normal input data, the algorithm is still save.

For example, if we are solving SAT, but the inputs take only the form of horn clauses (for example $\bigwedge x_i \to y$), then the problem is solvable in PTIME.

---

[15]So many in fact, that the author struggled to find any interesting PTIME-complete problems

It could also be, that the algorithm runs in $\mathcal{O}((1+\varepsilon)^n)$ for some small $\varepsilon$. Then $n$ would have to be very big for it to be problematic.

**Parametrization** The problem might be slow in general, but there are different parameters to the problem, that, when fixed, make the algorithm polynomial. Only increasing these parameters then give the full problem.

For example, checking if a given program in the ML language[16] does not violate any typing (i.e. a cast would be necessary) is `NPTIME`-complete[17], but if we parametrize by the number of type definitions in the system, we have merely a polynomial runtime.

**Approximation** As seen in section 3.3.3, many `NPTIME`-complete problems are actually optimization problems in disguise. Instead of looking for *the* optimal solution, often it is enough to settle for a sub-optimal, but good enough solution. In this text, we didn't develop the necessary tools to analyse such inaccurate algorithms, but such algorithms can come –provably– very close to an optimal solution, while taking only polynomial time.

### 3.3.4 Other complexity classes

The complexity classes introduced are of course only scratching the surface of the fascinating topic of complexity theory. There are many different modes of computation, that do not factor into `PTIME` or `NPTIME`, for example parallelism, which will be of ever greater importance, as the number of cores of computers increase, when the sequencial speed cannot. A bit further down the road, the quantum computer will get its share.

`NC` stands for Nick's class[18] and has been described as the class of well-parallelizable problems[19]. It is known that `NC` $\subset$ `PTIME`, but it is unknown, whether this inclusion is proper. Similarly to `NPTIME`-complete problems, `PTIME`-complete problems, such as CFG, are analyzed to prove this.

`BQP` is the class of bounded-error polynomial-time quantum computable decision problems. Bounded error compensates for the fact that quantum computation is, in order to achieve its full strength, random. Bounding the error means that it gives the wrong answer at most $0 < \varepsilon < \frac{1}{2}$ of the time, where $\varepsilon$ is often arbitrarily chosen to be $\frac{1}{3}$. Given a procedure in `BQP`, it is easy to see, that we can lower the probability of error arbitrarily low by running the procedure multiple times and taking the average answer.

It is possible to simulate a quantum computation in `PSPACE`, and it is possible to simulate a deterministic computation on a quantum computer, so `PTIME` $\subset$

---

[16]Java or C++ would be no better as anyone who ever used template code knows. Actually, C++ templates are even Turing-complete [14]

[17]See [2]

[18]honoring Nick Pippenger

[19]See [12]

BQP $\subset$ PSPACE, but again, all inclusions are suspected, but not proven, to be proper.

The connection to NPTIME is unknown, but NPTIME $\subset$ BQP would imply NPTIME $\neq$ PTIME.